%% file: paper.tex
\documentclass[screen, sigconf]{acmart} 
\definecolor{shmgreen}{RGB}{126, 166, 61}

\newcommand {\bjoern}[1]{}
\newcommand {\shm}[1]{}

\usepackage{array}

 \newcommand{\edits}[1]{#1}
 \newcommand{\deletes}[1]{}

\definecolor{OliveGreen}{HTML}{3C8031}
\definecolor{Fuchsia}{HTML}{8C368C}

\AtBeginDocument{%
  }

\copyrightyear{2026}
\acmYear{2026}
\setcopyright{cc}
\setcctype{by}
\acmConference[C\&C '26]{Creativity and Cognition}{July 13--16, 2026}{London, United Kingdom}
\acmBooktitle{Creativity and Cognition (C\&C '26), July 13--16, 2026, London, United Kingdom}
\acmDOI{10.1145/3803784.3807532}
\acmISBN{979-8-4007-2583-8/2026/07}

\begin{document}

\title{Artographer: a Curatorial Interface for Art Space Exploration}




\author{Shm Garanganao Almeda}
\authornote{Work done as a visiting researcher at Midjourney.}
\orcid{0000-0001-7660-313X}
\affiliation{%
  \institution{University of California, Berkeley}
  \city{Berkeley}
  \state{CA}
  \country{USA}
}

\author{John Joon Young Chung}
\orcid{0000-0002-8492-2525}
\affiliation{%
  \institution{Midjourney}
  \city{San Francisco}
  \state{CA}
  \country{USA}}

\author{Sophia Liu}
\orcid{0009-0008-7746-0749}
\affiliation{%
  \institution{University of California, Berkeley}
  \city{Berkeley}
  \state{CA}
  \country{USA}
}

\author{Yuwen Lu}
\orcid{0000-0003-0845-5563}
\authornotemark[1]
\affiliation{%
  \institution{University of Notre Dame}
  \city{Notre Dame}
  \state{IN}
  \country{USA}}

\author{Brett Halperin}
\authornotemark[1]
\orcid{0000-0002-3555-6637}
\affiliation{%
  \institution{University of Washington}
  \city{Seattle}
  \state{WA}
  \country{USA}
}

\author{Bjoern Hartmann}
\orcid{0000-0002-0693-0829}
\affiliation{%
  \institution{University of California, Berkeley}
  \city{Berkeley}
  \state{CA}
  \country{USA}
}

\author{Max Kreminski}
\orcid{0009-0002-6268-4033}
\affiliation{%
  \institution{Cornell Tech}
  \city{New York}
  \state{NY}
  \country{USA}}

\renewcommand{\shortauthors}{Almeda et al.}

\begin{abstract}
  \input{sections/0-abstract}
\end{abstract}

\ccsdesc[500]{Human-centered computing~Empirical studies in HCI}
\ccsdesc[500]{Human-centered computing~Interactive systems and tools}
\keywords{art distribution, media platform design, creativity support tools, creativity supportive ecosystems, art history}
\begin{teaserfigure}
  \includegraphics[width=\linewidth]{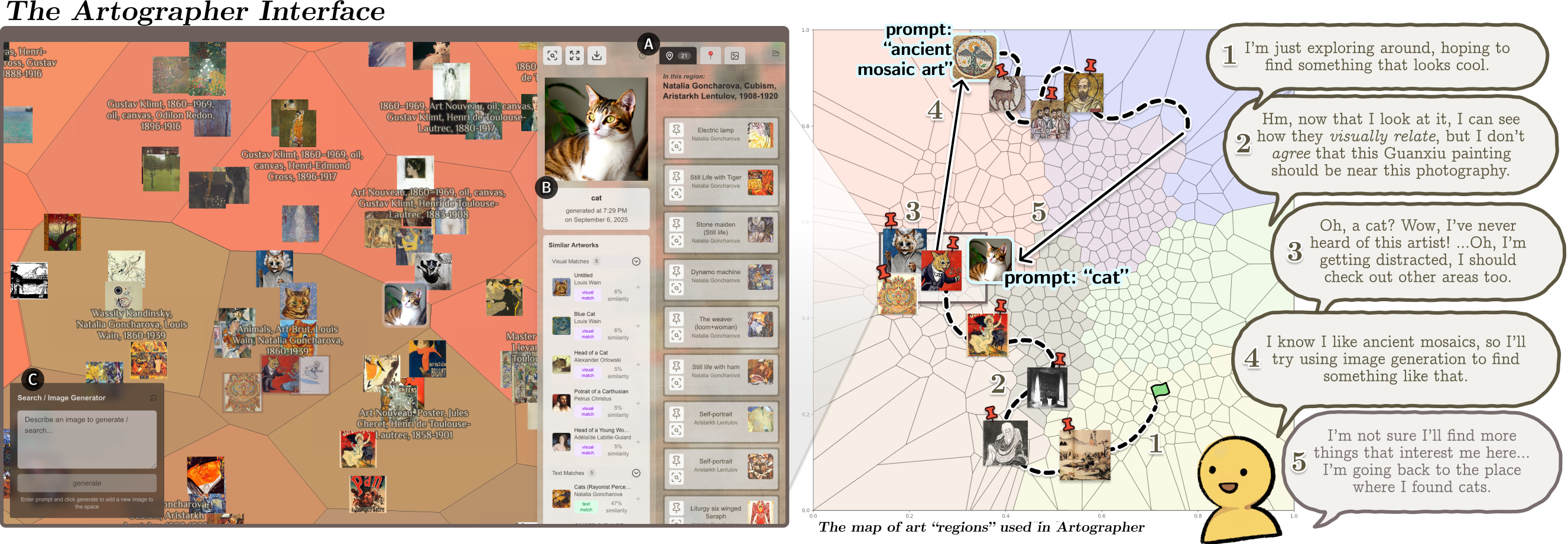}
  \caption{Artographer is an interface for exploring a curated dataset of \textasciitilde16,000 historical artworks as a zoomable, similarity-clustered map, constructed from multiple embedding models' representations of each artwork’s visual and semantic features. We traced how 20 participants, including 9 art history scholars, used Artographer to discover and collect across art space.}
  \Description{This figure shows the Artographer interface, a system for exploring approximately 16,000 artworks from the public domain through an interactive 2D zoomable map. The visualization organizes artworks based on multimodal embeddings that capture both visual features and semantic metadata. The left panel displays the main exploration interface where artworks appear as thumbnails distributed across a colored terrain map. The spatial arrangement reflects similarity relationships - artworks with related visual and semantic features cluster together. Labels scattered throughout identify specific artists, medium, or artwork titles. The left panel, A, shows specific artworks within a selected area. The center panel, B, shows images that is relevant to the generated image. The right-side panel, C, shows a panel for search or image generation. The right side of the figure demonstrates a user interaction scenario. A simulated user explores the collection with the following lines: 1. "I'm just exploring around, hoping to find something that looks cool." 2. "Hm, now that I look at it I can see how they visually relate, but I don't agree that this Guanxiu painting should be near this photography." 3. "Oh a cat? Wow, I've never heard of this artist! ... Oh I'm getting distracted, I should check out other areas too." 4. "I know I like ancient mosaics, so I'll try using image generation to find something like that." 5. "I'm not sure I'll find more things that interest me here... I'm going back to the place where I found cats." The interface shows how the system responds to prompts by highlighting relevant regions on the map and suggesting related artworks.}
  \label{fig:teaser}
\end{teaserfigure}


\maketitle

\input{sections/1-introduction}
\input{sections/2-related-works}

\input{sections/3-methods}
\input{sections/4-findings}

\input{sections/5-discussion}

\input{sections/6-future-work}
\input{sections/7-conclusion}



\bibliographystyle{ACM-Reference-Format}
\bibliography{sections/manual_references}

\newpage
\appendix


\onecolumn
\section{Appendix}
\setcounter{figure}{0}
\renewcommand\thefigure{A\arabic{figure}}
\setcounter{table}{0}
\renewcommand\thetable{A\arabic{table}}

\subsection{Participant Table}

\begin{table}[H]
\small
\caption{Participant Background Information and Study Conditions}
\label{tab:participants}
\begin{tabular}{p{0.5cm}p{3.2cm}p{3.2cm}p{3.5cm}>{\raggedright\arraybackslash}p{2.0cm}>{\raggedright\arraybackslash}p{1.8cm}}
\hline
\textbf{P\#} & \textbf{Experience as a Creative Practitioner} & \textbf{Experience Studying Art as a Subject} & \textbf{Art Study Background} & \textbf{System Order} & \textbf{Theme Order} \\
\hline
P01 & Significant experience & Moderate experience & experience taking Art History courses & Baseline, Artographer & Conflict, Peaceful \\
\hline
P02 & Expert experience & Expert experience & degree in Design; teaches Art \& Design courses & Artographer, Baseline & Conflict, Peaceful \\
\hline
P03 & Novice-level experience & Novice-level experience & experience taking Art Practice courses & Baseline, Artographer & Peaceful, Conflict \\
\hline
P04 & Moderate experience & Novice-level experience & experience taking Art Practice courses & Artographer, Baseline & Peaceful, Conflict \\
\hline
P05 & Novice-level experience & Significant experience & Art History graduate student researcher & Baseline, Artographer & Conflict, Peaceful \\
\hline
P06 & Moderate experience & Moderate experience & experience taking Art History courses & Artographer, Baseline & Conflict, Peaceful \\
\hline
P07 & Moderate experience & Little to no experience & experience taking Design courses & Baseline, Artographer & Peaceful, Conflict \\
\hline
P08 & Significant experience & Moderate experience & experience taking Design courses & Artographer, Baseline & Peaceful, Conflict \\
\hline
P09 & Significant experience & Significant experience & experience taking Art History courses, self-study & Baseline, Artographer & Conflict, Peaceful \\
\hline
P10 & Novice-level experience & Novice-level experience & experience taking Art History courses & Artographer, Baseline & Conflict, Peaceful \\
\hline
P11 & Moderate experience & Novice-level experience & self-study, visits art museums and gallery shows & Baseline, Artographer & Peaceful, Conflict \\
\hline
P12 & Expert experience & Expert experience & graduate degree in Art History & Artographer, Baseline & Peaceful, Conflict \\
\hline
P13 & Expert experience & Expert experience & degree in Art Practice, took courses in art history, experience working in an art museum & Baseline, Artographer & Conflict, Peaceful \\
\hline
P14 & Novice-level experience & Expert experience & degree in Art History, experience working in an art museum & Artographer, Baseline & Conflict, Peaceful \\
\hline
P15 & Expert experience & Moderate experience & degree in Art Practice, took courses in Art History & Baseline, Artographer & Peaceful, Conflict \\
\hline
P16 & Significant experience & Expert experience & Art History graduate student researcher & Artographer, Baseline & Peaceful, Conflict \\
\hline
P17 & Expert experience & Moderate experience & experience taking Art History courses, self-study to support creative practice & Baseline, Artographer & Conflict, Peaceful \\
\hline
P18 & Novice-level experience & Novice-level experience & experience taking Art History courses & Artographer, Baseline & Conflict, Peaceful \\
\hline
P19 & Significant experience & Significant experience & self-study to support creative practice & Baseline, Artographer & Peaceful, Conflict \\
\hline
P20 & Moderate experience & Expert experience & degree in Art History, Art History research experiences & Artographer, Baseline & Peaceful, Conflict \\
\hline
\end{tabular}
\end{table}

\twocolumn

\section{Post-Study Task Reflection Questionnaire}

\subsection{Task 1, Part 1}


\begin{enumerate}
    \setcounter{enumi}{0}
    
    \item \textbf{What was your target theme?}
    \begin{itemize}
        \item ``Peaceful''
        \item ``Fear''
        \item ``Happy''
        \item ``Sad''
        \item ``Conflict''
    \end{itemize}
\end{enumerate}

\noindent\textit{Rate your response to the following questions from 1 (Strongly Disagree) to 7 (Strongly Agree)}

\begin{enumerate}
    \setcounter{enumi}{2}
    
    \item \textbf{I feel confident that the images I collected express the target theme.} (Required) \\
    1 (Strongly Disagree) -- 2 -- 3 -- 4 -- 5 -- 6 -- 7 (Strongly Agree)
    
    \item \textbf{I feel confident that, upon viewing a selection of the images I submitted, others would be likely to correctly identify which of the 5 themes I was assigned.} (Required) \\
    1 (Strongly Disagree) -- 2 -- 3 -- 4 -- 5 -- 6 -- 7 (Strongly Agree)
    
    \item \textbf{What I was able to collect was worth the effort I had to exert to collect it.} (Required) \\
    1 (Strongly Disagree) -- 2 -- 3 -- 4 -- 5 -- 6 -- 7 (Strongly Agree)
    
    \item \textbf{It was easy for me to explore many different ideas, options, or outcomes, using this system.} (Required) \\
    1 (Strongly Disagree) -- 2 -- 3 -- 4 -- 5 -- 6 -- 7 (Strongly Agree)
    
    \item \textbf{I enjoyed using the system.} (Required) \\
    1 (Strongly Disagree) -- 2 -- 3 -- 4 -- 5 -- 6 -- 7 (Strongly Agree)
    
    \item \textbf{I was satisfied with what I got out of the system.} (Required) \\
    1 (Strongly Disagree) -- 2 -- 3 -- 4 -- 5 -- 6 -- 7 (Strongly Agree)
    
    \item \textbf{I would be happy to use this system on a regular basis.} (Required) \\
    1 (Strongly Disagree) -- 2 -- 3 -- 4 -- 5 -- 6 -- 7 (Strongly Agree)
    
    \item \textbf{The system was helpful in allowing me to track different ideas, outcomes, or possibilities.} (Required) \\
    1 (Strongly Disagree) -- 2 -- 3 -- 4 -- 5 -- 6 -- 7 (Strongly Agree)
    
    \item \textbf{Please feel invited to use this space to share any additional thoughts or comments you might have about your experience.} \\
    \textit{[Open response]}
\end{enumerate}

\noindent\textit{Thanks! Please pause here and let the interviewer know that you are ready to move on.}

\subsection{Task 1, Part 2}


\noindent\textit{Rate your response to the following questions from 1 (Strongly Disagree) to 7 (Strongly Agree)}

\begin{enumerate}
    \setcounter{enumi}{11}
    
    \item \textbf{What was your target theme?}
    \begin{itemize}
        \item ``Peaceful''
        \item ``Fear''
        \item ``Happy''
        \item ``Sad''
        \item ``Conflict''
    \end{itemize}
    
    \item \textbf{I feel confident that the images I collected express the target theme.} (Required) \\
    1 (Strongly Disagree) -- 2 -- 3 -- 4 -- 5 -- 6 -- 7 (Strongly Agree)
    
    \item \textbf{I feel confident that, upon viewing a selection of the images I submitted, others would be likely to correctly identify which of the 5 themes I was assigned.} (Required) \\
    1 (Strongly Disagree) -- 2 -- 3 -- 4 -- 5 -- 6 -- 7 (Strongly Agree)
    
    \item \textbf{What I was able to collect was worth the effort I had to exert to collect it.} (Required) \\
    1 (Strongly Disagree) -- 2 -- 3 -- 4 -- 5 -- 6 -- 7 (Strongly Agree)
    
    \item \textbf{It was easy for me to explore many different ideas, options, or outcomes, using this system.} (Required) \\
    1 (Strongly Disagree) -- 2 -- 3 -- 4 -- 5 -- 6 -- 7 (Strongly Agree)
    
    \item \textbf{I enjoyed using the system.} (Required) \\
    1 (Strongly Disagree) -- 2 -- 3 -- 4 -- 5 -- 6 -- 7 (Strongly Agree)
    
    \item \textbf{I was satisfied with what I got out of the system.} (Required) \\
    1 (Strongly Disagree) -- 2 -- 3 -- 4 -- 5 -- 6 -- 7 (Strongly Agree)
    
    \item \textbf{I would be happy to use this system on a regular basis.} (Required) \\
    1 (Strongly Disagree) -- 2 -- 3 -- 4 -- 5 -- 6 -- 7 (Strongly Agree)
    
    \item \textbf{The system was helpful in allowing me to track different ideas, outcomes, or possibilities.} (Required) \\
    1 (Strongly Disagree) -- 2 -- 3 -- 4 -- 5 -- 6 -- 7 (Strongly Agree)
    
    \item \textbf{Please feel invited to use this space to share any additional thoughts or comments you might have about your experience.} \\
    \textit{[Open response]}
\end{enumerate}

\noindent\textit{Thanks! Please pause here and let the interviewer know that you are ready to move on.}

\subsection{Freeform Exploration Task}

Here we ask you to respond to your experience using the system for Freeform Exploration.

\begin{enumerate}
    \setcounter{enumi}{21}
    
    \item \textbf{It was easy for me to explore many different ideas, options, or outcomes, using this system or tool.} (Required) \\
    1 (Strongly Disagree) -- 2 -- 3 -- 4 -- 5 -- 6 -- 7 (Strongly Agree)
    
    \item \textbf{I was satisfied with what I got out of the system or tool.} (Required) \\
    1 (Strongly Disagree) -- 2 -- 3 -- 4 -- 5 -- 6 -- 7 (Strongly Agree)
    
    \item \textbf{I enjoyed using the system.} (Required) \\
    1 (Strongly Disagree) -- 2 -- 3 -- 4 -- 5 -- 6 -- 7 (Strongly Agree)
    
    \item \textbf{The system or tool was helpful in allowing me to track different ideas, outcomes, or possibilities.} (Required) \\
    1 (Strongly Disagree) -- 2 -- 3 -- 4 -- 5 -- 6 -- 7 (Strongly Agree)
    
    \item \textbf{What I was able to collect was worth the effort I had to exert to collect it.} (Required) \\
    1 (Strongly Disagree) -- 2 -- 3 -- 4 -- 5 -- 6 -- 7 (Strongly Agree)
    
    \item \textbf{I would be happy to use this system on a regular basis.} (Required) \\
    1 (Strongly Disagree) -- 2 -- 3 -- 4 -- 5 -- 6 -- 7 (Strongly Agree)
    
    \item \textbf{Please submit your 3 favorite images from the study. Below, we'll ask you to write a brief reflection about each of the images you submit.} \\
    \textit{These can be historical artworks you found during the study, images you generated during the study, or both.} \\
    \textit{[File upload]}
    
    \item \textbf{For each of the 3 images you submitted, please write 1--2 sentences to answer each of the following questions:}
    \begin{itemize}
        \item How did you find or generate this image? (were you looking for it?)
        \item Why did you select it as one of your favorites?
    \end{itemize}
    \textit{[Open response]}
    
    \item \textbf{Please feel invited to use this space to share any additional thoughts or comments you might have about your experience.} \\
    \textit{[Open response]}
\end{enumerate}
\onecolumn
\section{Likert-Scale Responses}

\begin{figure}[H]
    \centering
    \includegraphics[width=0.9\textwidth]{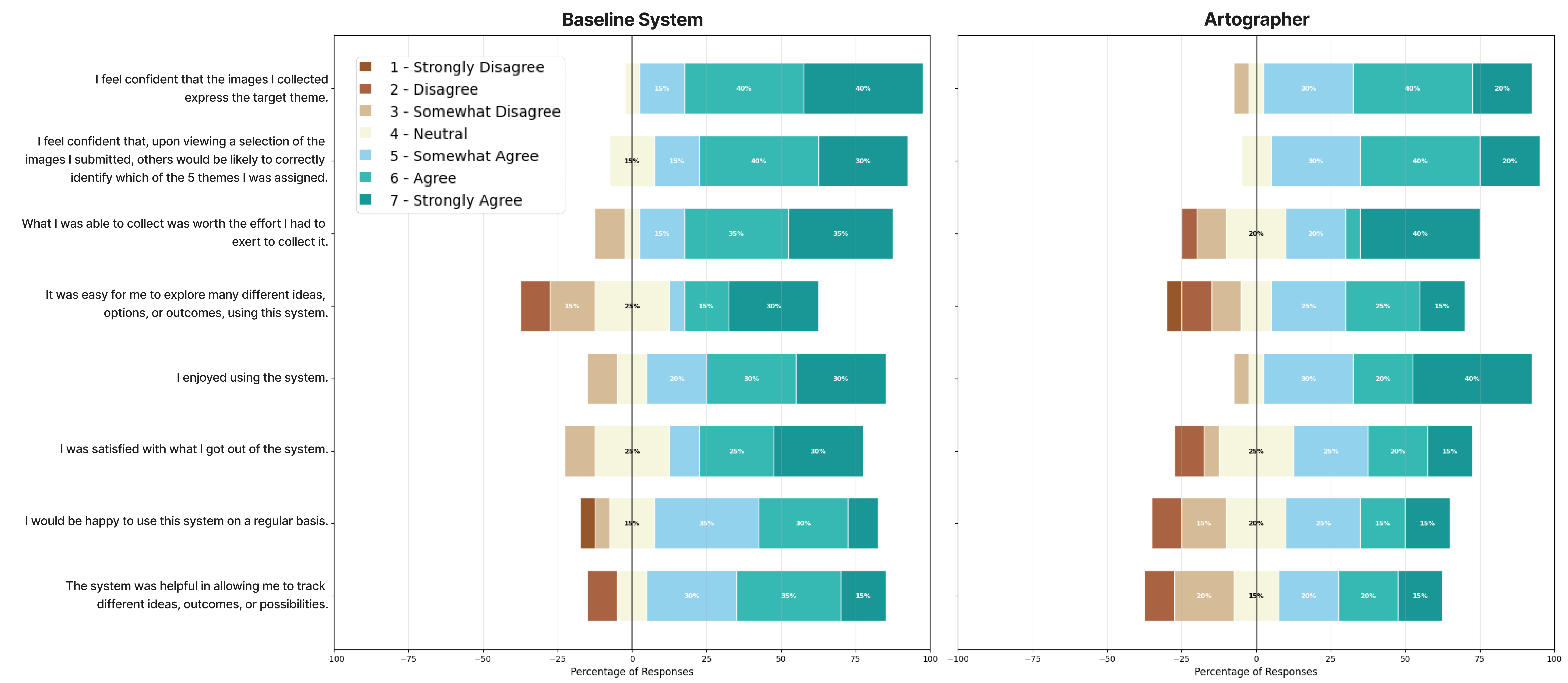}
    \caption{Participant Likert responses comparing the Baseline system and Artographer (brown = disagreement, blue–green = agreement). There is no statistically significant difference in responses between the two systems; participants were able to complete their exploration task effectively in both systems. }
    \Description{Two side-by-side horizontal stacked bar charts (Baseline on the left, Artographer on the right) showing distributions of responses to eight Likert statements:
    confidence that collected images express the target theme; confidence others would identify the assigned theme; worth the effort; ease of exploring ideas/options; enjoyment; satisfaction with outcomes; willingness to use regularly; and helpfulness for tracking ideas/outcomes. Bars run from disagreement on the left to agreement on the right (brown = disagreement, tan/neutral in the middle, blue–green = agreement). In both panels, agreement segments dominate—roughly about three-quarters of responses—with small disagreement segments. The two panels look very similar overall, indicating no consistent directional difference between systems.}
    \label{fig:likert-comparison}
\end{figure}

\end{document}

%% file: sections/0-abstract.tex
Relating a piece to previously established works is crucial in creating and engaging with art, but AI interfaces tend to obscure such relationships, rather than helping users explore them. Embedding models present new opportunities to support spatially exploring and relating artwork. We built Artographer, an art-exploration system featuring a zoomable 2-D map, constructed from similarity-clustered embeddings of \textasciitilde16,000 historical artworks. We used Artographer as a design probe to explore how alternative artwork distribution interface design can shape media engagement: we invited 20 participants, including 9 art history scholars, to traverse the map, collecting artworks for a goal-driven task and while freely exploring. We identify values enacted in spatial art discovery (Visibility, Agency, Serendipity, Friction) and consider how these values challenge dominant design paradigms---in particular, the recommendation systems governing contemporary media distribution platforms. We reimagine a \textit{curatorial} approach to media distribution, within digital ecosystems where history and culture can thrive.

%% file: sections/1-introduction.tex
\section{Introduction}

\noindent The world is inundated with images, at least some of which are art. 
Curation is the act of selecting and presenting artworks; \textit{how} those selection and presentation decisions are made crucially shapes the creative ecosystem around them---
the kinds of artistic support networks, and downstream cultural impacts, that result~\cite{almeda_creativity_2025, chung_artist_2022}.

While Creativity Support Tool (CST) research in HCI has studied and contributed to an increasingly vast space of tools for supporting the \textit{creation} of digital artifacts~\cite{frich_twenty_2018, frich_mapping_2019, chung_intersection_2021}, recent critical CST perspectives have identified an overemphasis on designing and evaluating CSTs for productivity and artifact-production~\cite{li_beyond_2023, rhys_cox_beyond_2025}. They call for methodological shifts towards \textit{artistic support} research, to acknowledge the socially and culturally entangled reality of art-making within creative media \textit{ecosystems}~\cite{chung_artist_2022, li_beyond_2023, almeda_creativity_2025, kato_power_2025}. Such work has recognized the role of creative \textit{distribution} and \textit{reception}---including the platforms available for share and engage with artwork---as central to the function of creativity supportive ecosystems where artistic communities can sustainably thrive~\cite{almeda_creativity_2025}. 

Yet, in contrast to the diverse multitude of CSTs accessible for \textit{creating} digital media, the research, design, and development of creative media distribution systems is increasingly dominated by a few private social media corporations~\cite{feng_mapping_2024}, to the detriment of user \textit{agency} in media interactions~\cite{lukoff_how_2021, baughan_i_2022, baumer_departing_2018}.

With user populations in the billions~\cite{dixon2025most}, social media platforms are establishing scrolling feeds of algorithmically recommended content as a dominant digital media interface design paradigm~\cite{widener2025digital}.  These systems enact the incentives of the corporations that design and govern them~\cite{von_davier_looking_2025} --- valuing forms of media ``engagement'' that center metrics like user \textit{retention}, advertising click-through, or frequency of actions that lead to purchases~\cite{pancha_pinnerformer_2022, chen_pinfm_2025, zou_reinforcement_2019}. As artists cede control over the distribution, visibility, and consumption of their work to these increasingly dominant, centralized distribution platforms, they are pressured to produce ``content'' aligned with embedded platform values~\cite{von_davier_looking_2025}.

 
 As large, opaque AI systems \textit{intermediate} more interactions with art and imagery than ever before \cite{von_davier_machine_2024},  widespread use of Generative AI (GenAI) is also rapidly accelerating the \textit{production} of images: text-to-image models were used to generate over 15 billion images in 2023 alone, exceeding the number of photographs produced in the first 150 years of photographic history \cite{valyaeva_ai_2023}. AI models excel at synthesizing and surfacing relationships across large bodies of multimodal data---yet mainstream AI-CSTs tend to present images \textit{detached} from the human contexts they materially derive from. 
 
This AI-driven media ecosystem is rapidly optimizing for undercontextualized content consumption that \textit{disincentivizes critical, reflective engagement with art and media}---threatening public media literacy and diversity of artistic expression~\cite{von_davier_looking_2025}.


How might the ways we select, present, and subsequently receive and \textit{engage} with media change when we embrace alternative values in the design of a creative distribution? Can we leverage the power of multimodal data-driven AI models to help users critically explore, reflect, and relate media, rather than obscuring these relationships?  

We ask:
\textbf{How might we reimagine media ``recommendation'' system design as ``curatorial'' system design?} That is, how can interfaces for selecting and presenting media embody the values of effective cultural art curation, shifting away from interaction design paradigms that optimize for ``content consumption'' toward \textit{facilitating personally meaningful, critically reflective engagements with art and media}?

\deletes{In this work, we use a frictional Research-through-Design (RtD) approach~\cite{pierce_tension_2021} to explore a new point in AI-supported media distribution design space that diverges from the monolithic ``content recommendation feed'' paradigm. To that end we design and develop \textit{Artographer}, a historical art exploration system. }
In this work, we \edits{design Artographer as a frictional alternative~\cite{pierce_tension_2021} to the ``content recommendation feed'' paradigm that dominates the design of AI media curation and presentation platforms.

Artographer is an art exploration system that} uses an intentionally opinionated, computational approach to curate a dataset of 15,958 historical artworks. It arranges this selection of artworks into a 2D similarity-clustered map, constructed from the combined vector spaces of multiple embedding models, and presented as an interactive, browser-based web interface. \edits{Artographer explores the use of \textit{contextualized AI generation} as a way of navigating this map: users can use text-to-image generation to add images to the map; these additions are placed in---and immediately relocate the user to---a neighborhood of similar historical artworks.}

\edits{We wanted to understand how computational curation can impact the way we encounter---and the ways we might \textit{design} future encounters---with media collections.  Rather than prescribing a particular design approach per se, we use Artographer as a research system to enact and investigate the implications of divergent media ``recommendation'' platform design.}

We recruited 20 creative community stakeholders (9 with significant art history or media curation expertise) to use Artographer, and to navigate and interrogate it as a computationally-constructed presentation of art. 
\edits{As a study instrument, Artographer allowed us to capture, trace, and characterize the ways users interacted with art within the interface. Their experiences also contextually grounded our interviews, and through} reflective, co-constructive discussions with participants, we collectively reconsider the future of computationally supported ``engagement'' with art and media. 

We report on exploration behaviors and experiences that characterize how participants explored and collected artworks in Artographer. We identify four curatorial values enacted in their spatial art explorations: Visibility, Agency, Serendipity, and Friction. We consider how these values can guide the design of media distribution systems that support deeper, more reflective~\cite{glinka_critical-reflective_2023, kreminski_reflective_2021} and dialectical~\cite{zhang_searching_2024}, engagements with art.

To sum, this work contributes...
\begin{enumerate}
    \item Artographer, an art exploration system that explores using...
    \begin{enumerate}
        \item an intentionally curated database of public domain historical artworks, featuring  \textasciitilde16,000 images with metadata,  \textasciitilde1,640 artist \& keyword text entries, and image, text, and multimodal embedding vectors for every artwork and keyword entry, and
        \item a zoomable, 2D visual-spatial map interface for exploring art space;\edits{
        \item contextualized image generation---a text-to-image generator that places AI-generated images in conversation with ``nearby'' historical artworks;
        }
    \end{enumerate}
    \item an empirical study conducted with n=20 participants, including 9 art history scholars, in which we instrument Artographer \edits{as a research system for tracing and characterizing how spatial map exploration can shape the reception of artwork, and elicit stakeholder perspectives on use of computational technologies in the curation and presentation of artwork;}
    \item \textit{Visibility}, \textit{Agency}, \textit{Serendipity}, and \textit{Friction} as four curatorial design values, towards reimagining media ecosystems where artistic communities and culture can thrive. 
\end{enumerate}

%% file: sections/2-related-works.tex
\section{Background \& Related Work}
Our work seeks to expand creativity support research into the design of media distribution platforms, towards alternative interfaces that support meaningful engagement with artwork. To this end, we leverage the well-established strengths of embedding models as multimodal relational systems, and zoomable 2-D maps as substrates for spatial-visual sensemaking. 
\subsection{Creative distribution support as a critical design space}
\textit{Artistic support research} has emphasized the importance of expanding HCI's understanding of ``creativity support'' beyond the task of designing tools to facilitate an individual artist's artifact creation process in isolation --- towards designing for the entangled realities of art-making in society~\cite{li_beyond_2023, chung_artist_2022, nakakoji_framework_1999, nakakoji_interaction_2002,  almeda_creativity_2025, kato_power_2025}.
Social theories of art describe art-making as a collective activity, performed by an \textit{art world}: a network of actors and sociotechnical systems performing interdependent, creativity supportive roles~\cite{becker_art_1974}.
\textit{Distribution} is central to art world activity; creative distribution systems determine which artists have the opportunity to thrive and influence the art world, versus those whose work remains unshared and unseen --- shaping media reception, and culture, in turn~\cite{almeda_creativity_2025}.

In traditional art worlds, humans perform creative distribution roles (e.g., publishers, curators), intermediating cultural engagement with art and media. While humans continue to perform these support roles on- and off-line, the work of media distribution is increasingly being delegated to computational systems.
With user populations in the billions, contemporary media consumption is increasingly dominated by social media platforms and streaming services that use algorithmic recommendation systems to determine what media is presented to users~\cite{von_davier_designing_2023}. 

Recommendation systems (RecSys) often center, and enact, the values of the corporations designing and governing them. For example, many are designed to optimize for \textit{engagement} and \textit{retention}-- presenting users with content similar to the content they have previously \textit{engaged} with, to encourage users to spend more time on the platform~\cite{pancha_pinnerformer_2022, chen_pinfm_2025}. 
The recommendation algorithms that govern curation on dominant media platforms pressure artists to conform to reductive embedded metrics, towards \textit{ecosystemic} issues --- e.g., creating threats to public media literacy, and to the \textit{diversity} of artistic expressions sustained by our media culture~\cite{von_davier_looking_2025}.

This motivates our interest to explore the design of divergent media distribution systems in tension with the dominant design progression trends~\cite{pierce_tension_2021}. In this work, we move to reimagine media ``recommendation'' system design as ``curatorial'' system design. How might the computational systems we design to support the selection, presentation, and reception of creative work diverge from the values of dominant media platform corporations? How  might they draw from the values of intentional artistic curation instead? Through the development and research instrumentation of Artographer, we seek to enact and investigate a point in media distribution design space---to explore how leveraging multimodal feature extraction, 2D spatial exploration, and other interaction design techniques for sensemaking rich cultural data can help us envision a future where computational ``recommendation'' systems support more meaningful engagements with art and media.
\edits{
\subsection{Computational interventions into art curation and presentation}
The design and development of Artographer builds upon a lineage of systems and techniques for the transformation and projection of cultural data collections, to support the expert research of art historians, curators, and digital humanities scholars~\cite{tuscher_nodes_2025, saleh_large-scale_2016}. 
Multimodal data models allow researchers to use visual and semantic traits to explore artwork collections, rather than textual searching across metadata alone; Replica~\cite{kaplan2016visual} used feature extraction to support using multimodal algebraic expressions to query large collections of paintings. Systems that incorporate dimensionality reduction techniques like UMAP and t-SNE help researchers visualize, analyze, and sensemake rich cultural collections~\cite{oygard_visualizing_2018, diagne_t-sne_2018, pietsch_cpietschvikus-viewer_2026}; notably, PixPlot, developed in 2017 by Yale's Digital Humanities Laboratory~\cite{leonard_pleonard212pix-plot_2026}, has been used to support a number of visual analysis studies (e.g.,~\cite{colwell2023distant}.) A recent and notable system in this space is Ohm et al.'s Collection Space Navigator~\cite{ohm_collection_2023}, a browser-based tool that combines 2D projections with configurable filters.
Configurable collection exploration tools offer critical support for expert art historians, researchers, and curators, allowing them to generate, query, filter and transform views on multifaceted collections of cultural data---to gain more comprehensive understanding or investigate specific research questions. Artographer draws on the techniques developed through this work---multimodal feature extraction, similarity clustering, dimensionality reduction, and interactive spatial projections---to explore the implications of incorporating the values of thoughtful and critical art curation into the design of a general media distribution interface.}

While using data science to support thoughtful \textit{expert} interpretation of historical artworks is fairly well-established, AI- and data-driven interventions into \textit{non-experts'} engagements with art are often designed to \textit{automate} interpretation. 
Computational interventions may position AI, implicitly or explicitly, as a ``solution'' to the ``subjectivity’’ of human interpretation---as a means to more efficiently replicate, automate, or replace the creative activities of human evaluators~\cite{glinka_critical-reflective_2023}. 
Examples include models that automate art-style classification \cite{luo_art_2025, bar_classification_2015, li_enhanced_2025} or predict visual ``aesthetic quality'' \cite{beaumont_laion-aiaesthetic-predictor_2022,  schuhmann_christophschuhmannimproved-aesthetic-predictor_2025}. Such classifiers can then be used to automatically curate end-user media recommendations, craft ``aesthetically'' curated datasets (e.g., by excluding samples that score below some measure of aesthetic quality), and to ultimately train Generative AI models to produce more ``aesthetically'' aligned outputs~\cite{murray_ava_2012, schuhmann_laion-aesthetics_2022, zhang_inkthetics_2020}. 

Generative AI models themselves draw from datasets of \textit{existing creative work}; these ``reference materials'' are latent in the artifacts that they generate \cite{carlini_extracting_2023}. 
Dependency on existing work is a core feature of human art production; several major theories of art define an artwork by its relationship to established art pieces and movements \cite{levinson_defining_1979, levinson_refining_1989}. Relating a piece of media to the space of previously established works is crucial in curating, creating, and engaging with art. 
AI models, as systems that excel at synthesizing and surfacing relationships across large bodies of data, are actually well-equipped to support relational art interactions. Yet, mainstream AI-CSTs tend to \textit{obscure} critical relational information, presenting images detached from the human contexts they materially derive from. 

Despite this, few AI-systems are designed to support non-experts in relating and meaningfully engaging with artworks~\cite{srinivasan_see_2024, glinka_critical-reflective_2023, almeda_creativity_2025}.  von Davier positions the work of designing for deeper appreciation in the digital presentation of artworks as a necessary counter to the ``contentification'' of media~\cite{von_davier_looking_2025}.

This motivates our interest in developing an experimental art interface that attempts to leverage the rich relational meaning-making afforded by multimodal AI models---\edits{and to understand how using these computational materials in the design of distribution platforms can shape more meaningful encounters with art and culture.} Artographer explores contextualized AI generation---relating GenAI images to established art space by placing AI-generated images in spatial context with visually or semantically related historical artworks. \edits{
As a research system, Artographer allowed us to bring community members into the investigation of GenAI as a design material for media exploration, and to elicit their perspectives on the cultural implications of using it in this way. 
}

\edits{
\subsection{Designing interactions with art and culture}
}

Artographer builds upon work by cultural institutions and HCI researchers studying how people engage with art and cultural artifacts \cite{ryokai_artistic_2015, villaespesa_museum_2019}, and how we might design systems to support deeper appreciation and engagement with art\cite{gorichanaz_engaging_2020, ciolfi_articulating_2016, von_davier_designing_2023}. This includes systems designed for the unique needs of specific cultural domains\cite{zhao_enhancing_2018, avgousti_enhancing_2024}.

Prior museum HCI work includes designing and studying the effects of novel applications, devices, and interactive or embodied experiences for museum visitors \cite{kortbek_communicating_2008, wakkary_situated_2007, weilenmann_instagram_2013, spence_seeing_2019, kobeisse_moving_2023, cameron_museum_2023, petrelli_phone_2018}.
Throughout this work, researchers and cultural institutions are notably shifting away from simply delivering educational content or suggesting ``factually-correct'' interpretations of artworks, instead asking questions like, ``How can we facilitate \textit{personal} engagement with artworks?'' and, ``How can we \textit{provoke} and \textit{empower} viewers to bring their own narratives into the interpretation of artworks?''

Digital interfaces can leverage the unique affordances of digital space for unique artwork engagement experiences, while supporting the democratization and accessibility of artwork collections through online distribution\cite{an_art_2024}.
Meinecke et al.~\cite{meinecke_towards_2022} and Meyer et al. \cite{meyer_algorithmic_2024} experiment with object-detection and similarity-clustering as materials for designing virtual museum collection interactions. \edits{Although such systems are inherently limited in the kinds of relationships they can suggest, viewers valued how a novel, interactive lens for engaging with artworks (e.g., the ability to group and explore a collection by shared subject matter)} could facilitate serendipitous discovery and reflection.

\edits{\subsubsection{Spatial interfaces for creative exploration}}
\label{sec:sensemaking}
2-dimensional, manipulable spaces are revisited throughout HCI and interaction design research to enable sensemaking and exploration of large-scale data. 
While reducing a rich dataset to two dimensions risks the loss of important information~\cite{jeon2025dimensionality}, it remains a widely adopted strategy for exploratory data sensemaking due to its power in leveraging humans' intuitive sense of spatial-visual perception. It is often combined with other interaction techniques to enable more flexible data exploration~\cite{yi2005dust, fass2000picturepiper}, like \textit{zooming}. Semantic zoomability allows users to use vertical movement to transition between different hierarchies of a data representation~\cite{bederson1994pad, lamping1995focus}. Advancements in AI for creativity support have afforded more flexible representations and visualizations of multimodal data ---
Zoomable 2D space has thus emerged as a promising substrate for designing AI-CSTs to support dynamic sensemaking and creative exploration, as in Sensescape~\cite{suh2023sensecape}, Luminate~\cite{luminate2024suh}, and Patchview~\cite{chung2024patchview}.

The power of zoomable, 2D-spatial map projections for relationally sensemaking multidimensional creative data is well-established in interaction design \edits{\textit{and} cultural analytics research}. Artographer leverages these strengths towards the design of a rich, yet navigable interface for engaging with a collection of artworks.

%% file: sections/3-methods.tex
\section{System Design}
Artographer\footnote{Live demos of Artographer, the Baseline system, the Map Generation API, and more, are available here: \href{https://artographer.snailbunny.site/}{https://artographer.snailbunny.site/}} is a system for exploring an intentionally curated set of historical artworks, presented as a similarity-clustered, zoomable 2D spatial map. 
We describe three key stages of system design: data curation (selecting and designing our representation of a rich set of artworks), map-making (constructing the similarity-clustered hierarchical map presentation of the artworks), and building the frontend 3D web interface. 

\subsection{Data Curation}

When curating a presentation that can capture a depth of meaningful relationships, not all data can take equal importance. For example, grouping artworks by artist name will produce a different arrangement than grouping by country, or century. What information can and should take precedence is dependent on the curator’s particular goals and priorities, and \textit{the content and distribution of the dataset they are curating from. }

We found that attempting to \textit{avoid} prioritizing \textit{any} relationships led to the construction of spatial maps where positional groupings frequently appeared meaningless or shallow. We designed an intentionally opinionated computational curation approach that is dependent on how much, and what kinds, of relationships are available in the dataset, prioritizing (1) visual similarity, then (2)\textit{ salient keyword similarity}, then (3) similarity of other text metadata. 

We constructed a historical artwork dataset of 26,886 artworks from WikiArt and Artsy.net.
We also gathered a keyword dataset of 1,643 art terms -- entries include every artist in the database, as well as descriptive keywords like \textit{``Art Nouveau''} and \textit{``Metallic''} from the Artsy Genome Project, Artsy.net's system of over 1,000 terms for categorizing artworks. \cite{artsy_artsy-art-genome-project_2025}. Each keyword maintains a list of the artworks it describes (i.e., is a tag for), and vice versa.

While several existing ``WikiArt'' datasets are readily available~\cite{wikiartorg_hugganwikiart_2025}, they are typically curated for training art-style classification models or similar, prioritizing collection size over depth. These datasets often lack much of the rich metadata and textual descriptions that are available for many artworks on WikiArt, and that would help construct a more richly relational map. 
This led us to curate our own WikiArt artworks dataset, and to augment entries with descriptive and contextual tags from The Art Genome Project. ML datasets also frequently exclude rights information---we strived to include only artworks in the public domain or under open licenses (e.g., Getty Trust Open Content Program, Creative Commons) and to ensure that images retained their licensing information.

\begin{figure}
    \centering
    \includegraphics[width=0.99\linewidth]{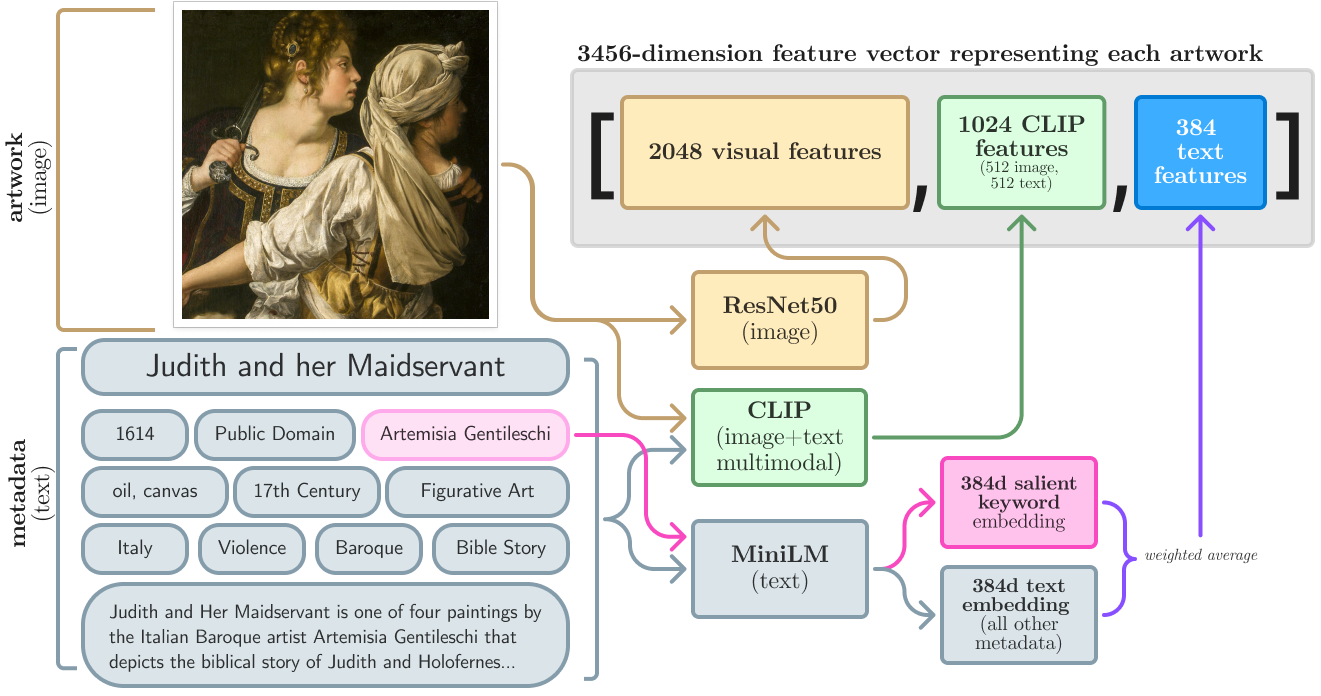}
    \caption{Each artwork is represented by a multimodal 3456d feature vector that combines visual embeddings generated with ResNet50, text embeddings from MiniLM, and image+text embeddings from CLIP. }
    \Description{This figure illustrates the multimodal embedding architecture used in Artographer to represent each artwork as a comprehensive 3456-dimensional feature vector. The system combines three distinct embedding approaches to capture both visual and semantic characteristics of artworks. The example shows Artemisia Gentileschi's "Judith and her Maidservant" from 1614, demonstrating how a single artwork is processed through multiple encoding pathways. The artwork image feeds into ResNet50 for visual feature extraction, producing 2048 visual features that capture compositional elements, color patterns, and stylistic attributes. The associated metadata, shown in gray boxes, includes textual information such as the title, artist name, date, medium (oil on canvas), period (17th Century), style (Figurative Art), location (Italy), thematic tags (Violence, Baroque, Bible Story), and a descriptive caption about the painting's subject matter. This metadata undergoes two parallel processing paths. One of them is colored in pink, indicating that this is a salient keyword. CLIP's multimodal encoder processes both the image and metadata text together, generating 1024 features that capture the semantic relationship between visual content and textual descriptions. Second, MiniLM processes the text independently, dealing with the salient keyword and other metadata separately. Then, embeddings for the salient keyword and other metadata are weight-averaged, creating 384 text-specific features that encode the semantic meaning of the metadata without visual context. The three embedding types—visual, multimodal, and textual—are concatenated into a single 3456-dimensional vector that comprehensively represents each artwork. }
    \label{fig:representing_artwork}
\end{figure}

We compute a high-dimensional representation of every artwork and keyword entry (See Figure \ref{fig:representing_artwork} for details). We combined visual embeddings (using artwork image data, via ResNet-50\footnote{https://huggingface.co/microsoft/resnet-50}), text embeddings (using artwork metadata, via MiniLM\footnote{https://huggingface.co/sentence-transformers/all-MiniLM-L6-v2}), and visual-text embeddings (using both image and metadata, via CLIP\footnote{https://huggingface.co/openai/clip-vit-base-patch32}) to ultimately represent each artwork with a 3,456-dimensional multimodal feature vector, where each kind of embedding is weighted with a manipulable parameter, to support dynamic adjustment. 

This strategy aligns with prior work that found a fused approach to be more effective at capturing hidden semantic relationships in artwork than using image or text embeddings alone~\cite{yilma_elements_2023}. 

In selecting metadata to use for each artwork's text embeddings, we chose a process that prioritizes certain \textit{salient keywords} over others, to intentionally bias the final presentation towards the more meaningful artwork groupings available for representation using our particular dataset. 

For each keyword, we calculate a salience score equal to (the number of artworks the keyword describes) * (the fraction of total artworks it describes):

\[
\text{salience} = \text{count} \times \frac{\text{count}}{\text{total\_artworks}}
\]

This deprioritizes ``niche'' keywords (that only describe a few pieces in the dataset) and overly broad keywords (like ``painting'', which describes such a large fraction of the dataset, it ceases to be meaningfully descriptive). 

We select a representative set of 500 salient keywords using a greedy iterative process: starting with the most salient keyword, we repeatedly add the next most salient keyword that describes \textit{the largest number of previously uncovered artworks}. This favors terms that provide distinctly descriptive relational information, rather than repeatedly characterizing groups of artworks already well-described by a more salient keyword. 

We used the set of artworks covered by these 500 salient keywords to reduce the dataset from 26,886 artworks, to the final set of 15,958 artworks. When computing each artwork's text embedding, we first take the embedding representing the salient keyword, then average it with an embedding that represents the remaining metadata as text. 
The result is a computational selection of artworks that curates for \textit{relationality}: all entries are described by at least one contextual term that relates it to a group of other artworks in the set. 

\subsection{Spatial Map-making Approach}
Our data curation strategy formed a strong set of base materials for a similarity-clustered presentation of artworks that embeds a variety of meaningful relationships. 

We use Uniform Manifold Approximation and Projection (UMAP)~\cite{mcinnes_umap_2018} to project the 3456-dimensional representation of each artwork onto 2-D space. 
Compared to t-SNE (another common dimensionality reduction technique), UMAP allows visualization designers to balance between local structure (data is well clustered within categories) and global structure (similar categories are co-located), and provides parameters for manipulating this balance \cite{coenen_understanding_2025}.

We first use UMAP on the entire dataset to construct a global structure, where groups of related artworks are clustered together.
We use k-means clustering to create a Voronoi diagram that contains each cluster within a cell---this forms Artographer's ``regional map.'' We iteratively merge these adjacent ``neighborhood''-like regions to create the larger, ``country''-like colored regions seen when viewing the map at a distance. 
We then re-run UMAP on each cluster, recomputing new 2D coordinates for each artwork, while containing them to the shape of their ``region.'' This achieves more meaningful local structure of the data \textit{within} each region when users view the map at a short distance. 

Some ``outlier artworks'' are so visually and semantically distinct from the rest of the dataset, they would be positioned too far to comfortably display or navigate to; we chose a reasonable map size, then iteratively nudge these outliers into the constraints of the map. (We report on interesting tensions and opportunities that arise from this presentation of ``marginalia'' in \S \ref{sec:outliers}.)

We developed a Flask Server API
that can perform each step of the map-making process on demand. The API returns a hierarchical 2D-map representation of the dataset as a nested JSON object, with 2D coordinates for each artwork, and supports changing a number of the variables involved in the map-making process---including the weights that adjust the ``strength'' of each kind of embedding.
The system supports generating a plurality of art-space maps, each capable of emphasizing different design priorities and ways of relating the data. For example, one can set the weight of text embeddings to zero, producing a map clustered by visual similarity alone.



\subsection{Frontend Web Interface}

\begin{figure*}
    \centering
    \includegraphics[width=0.98\linewidth]{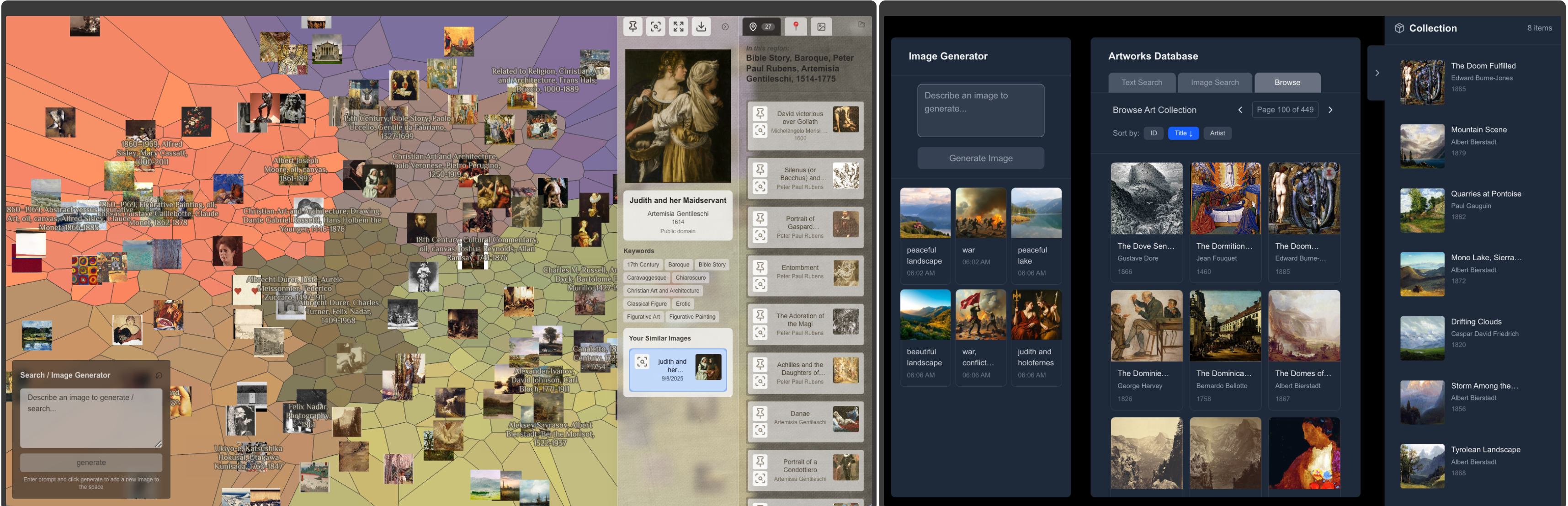}
    \caption{Artographer (left) is a spatial map interface for historical artwork exploration. Our Baseline system (right) was a Query-Based Search Interface that mimics traditional interfaces for searching an artwork database, with support for browsing, text-search (including approximate matches), and search-by-image (supported by an image generator.) Both interfaces support exploration of the same database of 15,958 artworks. }
    \Description{This figure compares two interfaces for exploring a database of 15,958 historical artworks: Artographer's spatial map approach (left) versus a traditional query-based search interface (right). The left panel shows Artographer's main exploration interface. Artworks appear as thumbnails distributed across a topographical map with colored regions transitioning from orange through purple to brown. The spatial arrangement reflects similarity relationships computed from the multimodal embeddings. Labels throughout identify artists or, and specific works. A detailed view popup shows "Judith and her Maidservant" by Artemisia Gentileschi with metadata including keywords, similar images panel, and an image generator search box at the bottom right. The right panel displays the baseline query-based interface with a dark theme. The top-left section contains an Image Generator prompt field. The middle has an Artworks Database with text and image search capabilities and a browse function. For now, the main browsing area shows a grid layout of artwork thumbnails with titles and metadata. A Collection panel on the right side shows saved artworks.}
    \label{fig:systems}
\end{figure*}

Users interact with the Artographer frontend interface through their web browser. The colorful ``regional map'' (see Fig.\ref{fig:systems} left) is drawn as a 2-D plane in a 3-D scene, rendered in Three.js using React-Three-Fiber. Each artwork is itself a 2-D plane ``floating'' on the map. Users click-and-drag, or use WASD/Arrow Keys, to pan. 

The user can scroll to ``zoom'' in, moving the scene's camera closer to the map. We limit the number of artworks that can be rendered at any time (presenting all 16k artworks would be cognitively and technically overwhelming). At high levels of zoom, we show only ``representative'' artworks. For each region, we choose the artwork closest to its centroid to ``represent'' it; we deterministically curate this list of representatives to a selection that is well-spread out across the current view of the map. As the user zooms in, the system renders more artworks into their viewport. 

The ``Selected Region'' tab (See ``A'' in Figure \ref{fig:teaser}) lists all artworks in the currently selected region, regardless of zoom level. Each artwork in this menu has two buttons: the ``Focus'' button pans the camera to center and zoom into that artwork on the map. The ``Pin'' button allows users to collect the artwork, adding it to their ``Pins'' menu. Pinning also makes the artwork visible at all times regardless of zoom level, allowing users to create landmarks.

\subsubsection{Image Generation as a Navigation into Established Art Space}
Users are provided a text-to-image generator, in the bottom-left corner of the interface labeled ``Search / Image Generator'' (See ``C'' in Figure \ref{fig:teaser}. 
When the user enters a prompt, Artographer makes an API request to a text-to-image model\footnote{https://huggingface.co/black-forest-labs/FLUX.1-schnell} --- as soon as the result is available, the system sends the image + prompt to the backend server. The system then extracts multimodal embeddings for the generated image + prompt, and finds its nearest neighbors in historical art space. Finally, the Artographer interface places the generated image in the 2D map, close to its \textit{visual} nearest neighbors. Generated images also provides ``links'' that allow the user to quickly navigate to/from the AI image and its nearest visual and semantic neighbors (B in Figure \ref{fig:teaser}) This positions image generation as a navigation into established art space, placing GenAI's outputs in conversation with historical, human-made artworks. For example, generating with the prompt,``bowl of fruit'' will move the user to the Still Life area of the map, and place the generated ``bowl of fruit'' image there as a reusable navigation landmark.  

\subsection{System Design Positionality \& Limitations}
The first author and lead system designer is a queer, Filipino-American artist and technology researcher whose experiences and communities are carried into this work. They wish to acknowledge that the public domain art collections that Artographer centers implicates this system design in the perpetuation of a Western-dominated art perspective---one that tends to overrepresent White, male, European voices. We lend from bell hooks in our approach, appreciating and utilizing mainstream artworks that resonate and serve the objective of this work, while remaining critical of ``the institutional frameworks through which work by this group is more valued than that of any other group of people in this society.''\cite{hooks_art_1995}

\section{Method}
 This study seeks to contribute to a methodological shift in CST research and design, away from developing and evaluating CSTs as teleological prototypes of progressional ``effectiveness'' and productivity~\cite{pierce_tension_2021, rhys_cox_beyond_2025, li_beyond_2023} for an \edits{isolated, monolithic user. We approach CST research as an opportunity to better understand creativity as supported by sociocultural networks, where various stakeholders play interdependent roles---and to recognize how new technology systems can both support and disrupt these ecosystems~\cite{chung_intersection_2021, palani_evolving_2024, almeda_creativity_2025, chung_artist_2022}.} 


\edits{Rather than seeking to design the ideal media-presentation system \textit{per se}, we sought to investigate a point in media presentation design space, and explore its implications.
To that end, we developed Artographer as a kind of trail-aware~\cite{hammad_towards_2025} cultural probe~\cite{boehner_how_2007} for tracing the creative activity~\cite{kreminski_herding_2026, hammad_tracing_2026} of artwork exploration, and for grounding co-constructive interviews with various stakeholders relevant to the design of media presentation and curation systems.}

\edits{We sought to use Artographer as a research system to investigate---and invite creative community members to participate in investigating---}the alternative normative ground it constructs and the counter-hegemonic design insights that emerge from it~\cite{li_beyond_2023}.
\subsection{Study Design}
We conducted a within-subjects study over video call, where all participants used both the Artographer spatial map interface and a Baseline system while sharing their screen. Our study design allowed both the interviewer and the participant to observe \textit{how} they navigated art space, grounding co-constructive discussions about their exploration experiences throughout. 

Each participant answered brief introductory questions about their background art experiences, then completed two targeted exploration tasks and a free exploration task, followed by an interview where they reflected on their experiences. 

Before each task, we sent participants a web link to access the Artographer or Baseline system. 
After each task, participants completed a survey form that asked Likert scale questions (drawn from \cite{louie_expressive_2022} and \cite{cherry_quantifying_2014}) and open-ended reflective questions about their experiences. Participants consented to the data collection and use, and were compensated with a \$50 gift card for \textasciitilde1 hour of their time. 

\subsubsection{Recruitment}
Altogether, we recruited 20 participants, including 9 with significant or expert experience in the study of art history. 
\edits{In line with a creativity supportive ecosystem minded approach, we sought a range of perspectives from stakeholders in various roles relevant to media distribution and engagement~\cite{chung_artist_2022, almeda_creativity_2025}.}

\edits{We distributed a call for ``art \& art history enthusiasts'' to participate in an ``art exploration study'' via social media, with a link to a screener survey that asked respondents to self-report their experience with art-making, and studying art as a subject---while emphasizing that novices with strong interest in art would be eligible to participate.\footnote{In line with institutional and organizational research protocols, we also required that participants be 18 years old or older and based in the United States.}} \edits{To recruit a wider range of expertise, we emailed a call for participation to art \& media history professors and graduate students from various institutions.}
 Experts' relevant background experiences included taking and teaching art history courses, conducting historical art research, and working in a museum; see our Appendix for participant details.
 
We selected from respondents to capture a range of experience levels and roles in the study, recruiting creative practitioners and art history experts, as well as ``novices'' with enthusiasm for consuming and engaging with artwork\edits{---for example, P11 frequents museums and art shows. See Table \ref{tab:participants} in the Appendix for participant details.}

\subsubsection{Targeted Exploration Tasks}
For each of two \textbf{targeted exploration tasks}, participants had 6 minutes to assemble a collection of images to express a particular theme. We asked them to:
\begin{enumerate}
    \item collect as many images as possible,
    \item capture multiple interpretations of the theme, and
    \item create a collection that could serve as a successful \textit{expressive communication} of their assigned theme. 
\end{enumerate}
Participants were told to imagine showing their final image collection to a third-party individual, who should then be able to guess which theme they had been assigned from five possible choices: \textit{Happy, Sad, Peaceful, Conflict, Fear}. Inspired by Louie et al.'s Expressive Communication evaluation \cite{louie_expressive_2022}, this task focused each participant on a common goal, while affording the kind of interpretive flexibility that characterizes creative activities. 

Participants completed the targeted task for each of two themes (``Peaceful'' and ``Conflict''), completing one task with Artographer and the other with a Baseline system.

Our use of a Baseline system was not intended to facilitate collection of directly comparable task-completion data, per se. Rather, completing a task with a Baseline interface provided participants a shared point of reference for co-constructive discussion. Artographer enacts an alternative design that diverges from dominant media interface paradigms; having a point for comparison helped researchers and participants ground our analysis of \textit{how} system design can shape engagement with artwork.

The Baseline system's design (see Figure \ref{fig:systems}) recreates common Query-Based Search interfaces used for art databases, with browsing, text-search (including semantic matching), and search-by-image (supported by an image generator.) Participants described that it felt straightforward and familiar; some compared it to interfaces they use to search museum catalogs and other digital cultural archives. Both interfaces supported exploration of the same database of 15,958 artworks.

Meanwhile, the version of Artographer used in this study intentionally limits traditional query-based searching affordances, allowing us to better study the exploration behaviors of interest to our research questions (namely, spatial map navigation, AI image generation as a tool to augment navigation).

We counterbalanced system-order and theme order, evenly distributing participants across the four possible system-theme combinations (see the Participant Table in the Appendix for details.) 

\subsubsection{Free Exploration Task}
For the \textbf{free exploration task}, participants had 8 minutes to freely explore. They could generate and collect as many artworks as they’d like, and explore as much or as little of the map as they were interested in. At the end of the task time, they were asked to pick their top 3 favorite images (historical or generated) and submit them to a survey form. The form then asked participants to write a brief reflective journal entry explaining how they found and selected each image. 

\subsubsection{Post-Task Interview}
After the free exploration task, we conducted a brief semi-structured interview asking participants to reflect on their experiences in both systems, and to help us understand what uniquely characterizes using a spatial-map interface for art exploration. 

\subsection{Data Collection and Analysis}
We used Zoom to record and transcribe participants' interviews, including their think-aloud processes while completing the tasks and screensharing. We logged every artwork that participants interacted with in both systems, as well as coordinates for every artwork interaction and every panning or zooming movement in the spatial map interface. 

We used this timed coordinate data to generate exploration trajectory figures capturing each participant's behavior during each task. We used a collaborative digital whiteboarding platform to create a board for each participant that combined their exploration behavior figures, interview transcripts, artworks collected for each task, and responses to the post-task survey questions. 

Two authors led the qualitative analysis of these materials using digital post-its to annotate. We take a reflexive, inductive, constructivist approach to the analysis of our qualitative data---where both researcher and participant have a role in constructing understanding, and where themes emerge from the creative labor of qualitative coding~\cite{braun_reflecting_2019,  charmaz_constructing_2006}.
We first identified patterns in participants’ exploration data, then used these to craft provisional codes~\cite{saldana_coding_2015} for our analysis of their think-aloud transcripts, interview responses, and open-ended survey responses. We clustered post-its to synthesize across these data sources, forming the reflective understanding of the themes that surfaced across participants' experiences that guides the findings we report and discuss below. 

%% file: sections/4-findings.tex
\section{Findings}
First, we report on creative exploration behaviors and participant reflections that characterize how participants used a spatial map interface to explore artworks. 
\subsection{Characterizing Exploration in Artographer}
\begin{figure}
\centering\includegraphics[width=0.99\linewidth]{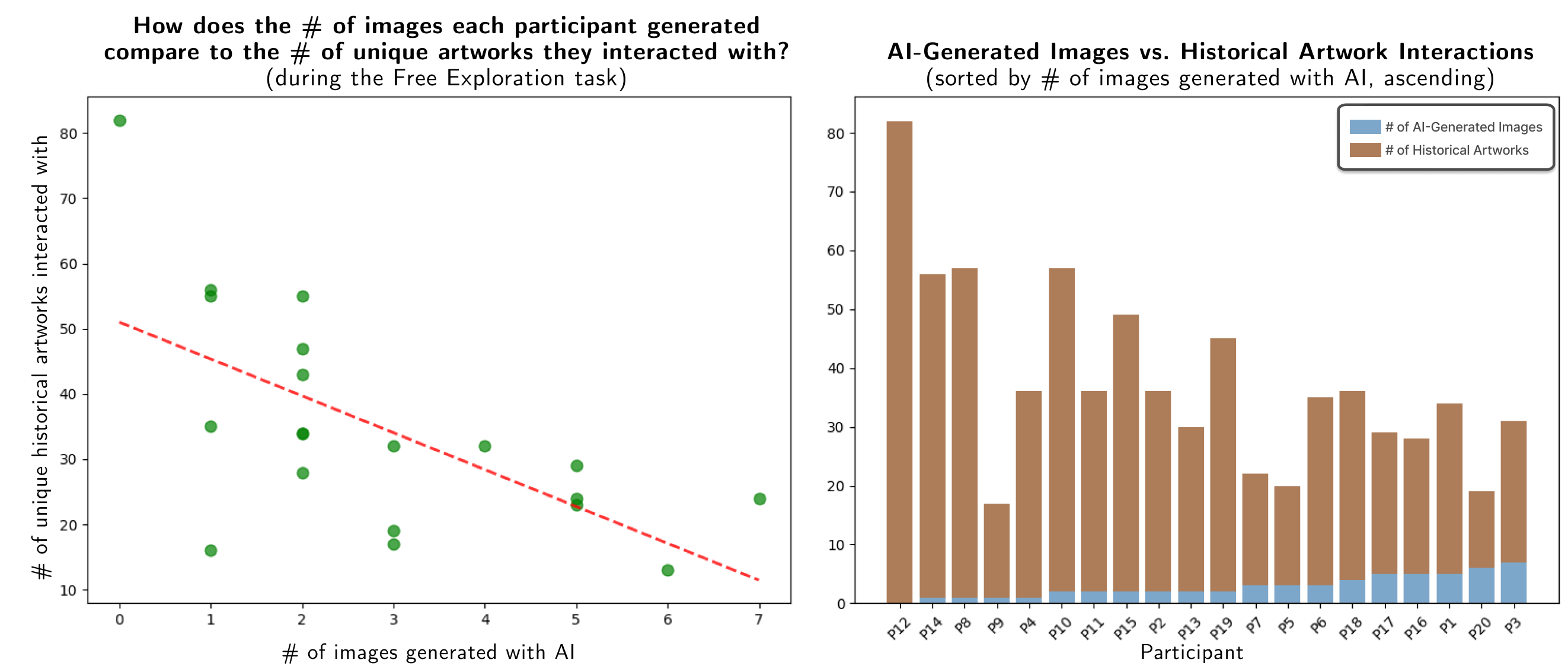}
    \caption{Comparing the number of images each participant generated with the number of unique artworks they interacted with. Participants who used wandering, rather than AI-generated jumping, explored more gradually, interacting with more images along the way.}
    \Description{This figure examines the relationship between AI image generation usage and historical artwork exploration during the free exploration task, revealing distinct exploration strategies among participants. The scatter plot on the left shows the correlation between number of AI-generated images (x-axis) and unique historical artworks interacted with (y-axis). A clear negative correlation emerges, indicated by the red dashed trend line. Participants who generated no AI images explored the most historical artworks, with 82 unique interactions. As AI generation increases to 5-7 images, historical artwork interactions drop to between 13 and 29. The stacked bar chart on the right presents individual participant data sorted by AI image generation frequency. Each bar shows blue segments for AI-generated images (bottom) and brown segments for historical artwork interactions (top). Participants P12 generated no AI images but explored 82 historical artworks, demonstrating extensive wandering behavior. As going left, the number of AI-generated images increases and there is a tendency that the number of interacted historical artworks decreases.}
    \label{fig:manual_vs_generating}
\end{figure}
 We found no statistically significant differences in responses to post-task Likert scale questions or task \textit{performance} when using the Baseline system vs. Artographer. As noted in our Method section, having a Baseline experience helped participants articulate how their \textit{experience} with the artworks differed between systems:
\begin{quote}
    ``The [Baseline] interface was actually way better for the targeted task, but the [Artographer]\footnote{participants referred to Artographer as ``the first system'' or ``the second system'' depending on system-order condition; for clarity, we replace these instances with the system name.} interface was way better for exploration.'' (P19)
\end{quote}
\begin{quote}
   ``I completed the task more efficiently with the [Baseline] interface, but I think this one was just more \textit{fun}.'' (P6)
\end{quote}
11/20 participants expressed a similar sentiment --- that Artographer was \textit{not} as easy or efficient to use as the Baseline system for a goal-driven task --- but it \textit{was} more \textit{fun}, especially for exploration. Our findings that characterize the ways a spatial map interface can impact \textit{how} participants explore provide insight into this. 


\subsection{Creative Exploration Behaviors}
\begin{figure}
    \centering
    \includegraphics[width=0.74\linewidth]{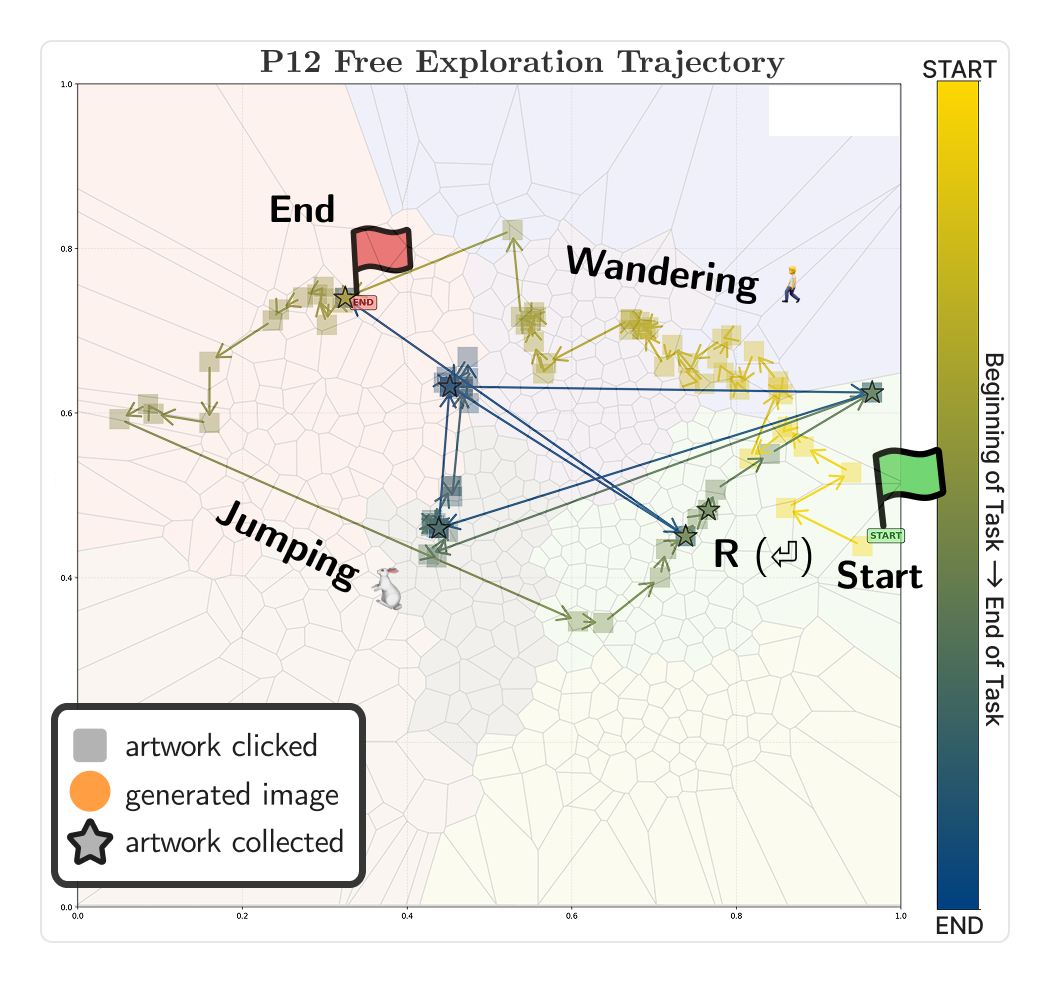}
    \caption{We logged the 2D coordinates of every event where a participant interacted with an image in the spatial map, and use these to visualize their exploration trajectory. P12 chose not to generate any images during their free exploration task --- they instead explored via a great deal of \textit{wandering} with some ``manual'' \textit{jumping} around the map.}
    \Description{This figure visualizes the exploration trajectory of Participant 12 during the free exploration task, demonstrating a "wandering" strategy without AI-generated image use. The trajectory is plotted on a 2D coordinate system representing the spatial map, with colors transitioning from yellow (start) to blue (end) indicating temporal progression. The exploration begins at the green "Start" flag marker in the lower right quadrant. The participant initially moves leftward through what appears to be a wandering phase. The dense cluster of yellow-to-green points connected by numerous short segments indicates the participant systematically explored a concentrated area, gradually shifting to the left side, examining many nearby artworks in sequence. This wandering pattern contrasts sharply with the afterward jumping movements, where the participant drastically changes their positions within the map. After a series of jumps, the exploration concludes at the red "End" marker in the upper left area where the participant used to wander around. Gray squares throughout the trajectory indicate clicked artworks, while stars mark collected pieces, showing that the participant was selective in their collection despite examining many options.}
    \label{fig:wandering}
\end{figure}
\begin{figure}
    \centering
    \includegraphics[width=0.74\linewidth]{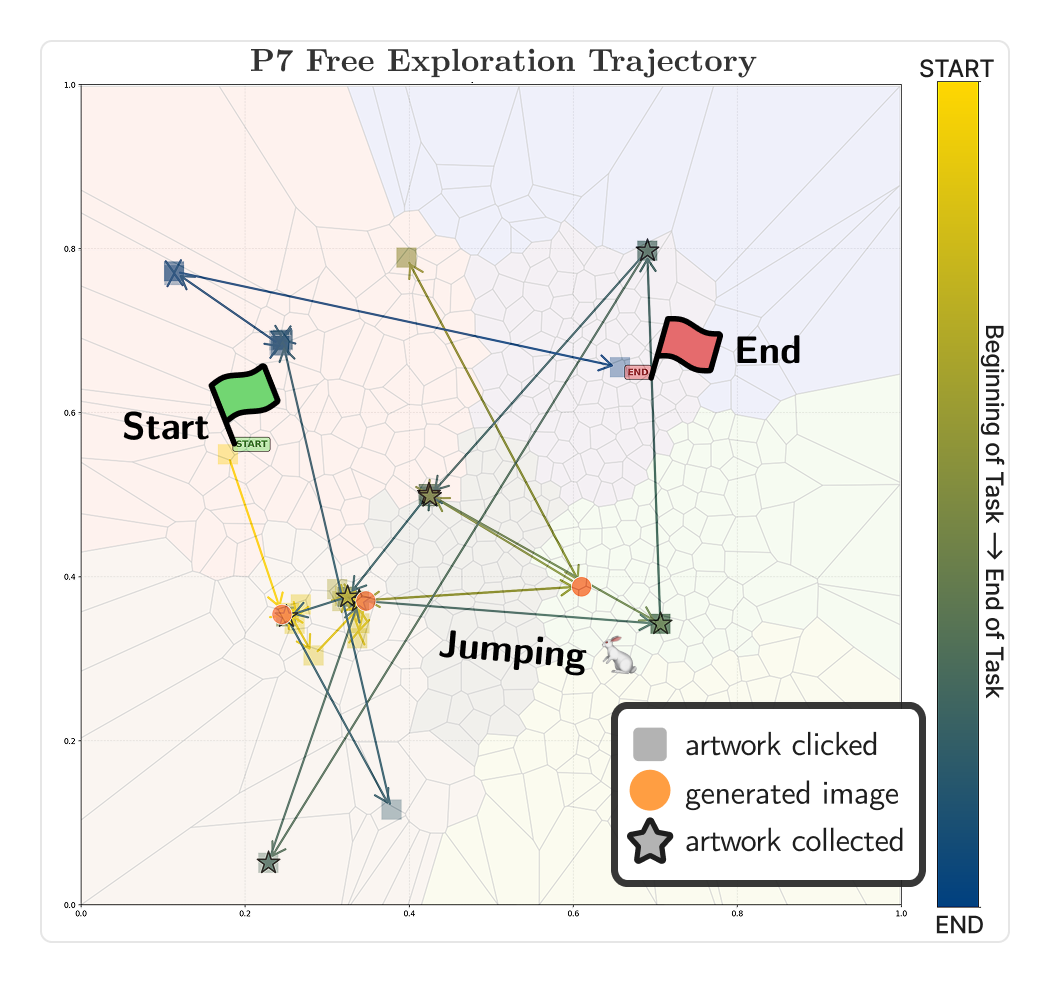}
    \caption{P7's exploration trajectory during the targeted collection task shows \textit{jumping} to search broadly, then deeper exploration (e.g. a cluster of artworks clicked and collected nearby) once a promising area is scouted.}
    \Description{This figure demonstrates Participant 7's exploration trajectory during the targeted collection task, illustrating a "jumping" strategy. The trajectory uses color coding from yellow (start) to blue (end) to show temporal progression across the 2D spatial map. Starting from the green marker in the upper-left region, P7 immediately employs a jumping strategy, making large movements across the map to scout different areas. The long connecting lines between distant points indicate deliberate navigation to sample diverse regions of the collection rather than exploring contiguously. This creates a web-like pattern centered roughly in the middle of the map. A notable orange dot appears in the lower-center area, marking where P7 generated an AI image. This generation serves as a strategic navigation tool, likely helping the participant reach a specific thematic region. The AI-generated image acts as a teleportation point to a new area of interest. The exploration concludes at the red "End" marker in the upper-right area after what appears to be a final targeted jump.}
    \label{fig:placeholder}
\end{figure}
\begin{figure*}
    \centering
    \includegraphics[width=0.99\linewidth]{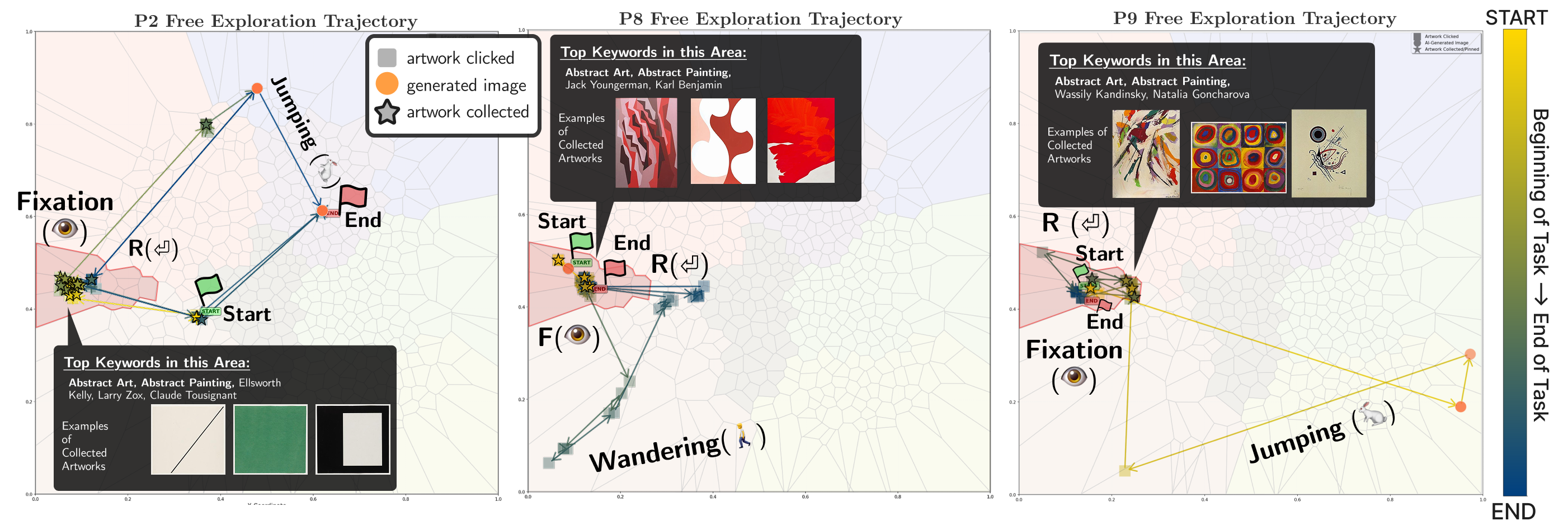}
    \caption{P3, P8, and P9 each described themselves as personally interested in graphic design and abstract art. During the free exploration task, they fixated on the area of the map that contained abstract art and painting, revisiting it after jumping to other areas. }
    \Description{This figure reveals how participants with shared interests in graphic design and abstract art exhibited similar exploration patterns, demonstrating the map's ability to support personal aesthetic preferences through spatial organization. P2's trajectory (left) begins at the green start marker and shows initial jumping movements before settling into a concentrated exploration of one area. The popup shows this region contains Abstract Art, Abstract Painting, and works by artists like Ellsworth Kelly, Larry Zox, and Claude Tousignant. The eye icon indicates fixation—repeated returns to this area after exploring elsewhere. P2 ends their exploration in the upper portion after circling back through the abstract art region. P8's trajectory (center) demonstrates initial exploration in the abstract art area, then wandering towards the lower portion of the map before jumping upward. The popup reveals this abstract art region contains Abstract Art, Abstract Painting, and works by Jack Youngerman and Karl Benjamin. P9's trajectory (right) shows extensive jumping across the map, indicated by long yellow lines connecting distant points. Despite this broad exploration strategy, P9 also fixates on the abstract art region, with the popup showing Abstract Art, Abstract Painting by Wassily Kandinsky and Natalia Goncharova.}
    \label{fig:designers}
\end{figure*}
In this section we report on how participants engaged with artwork in Artographer. We report the divergent and convergent behaviors identified in participants' exploration trajectories, contextualized alongside qualitative analysis of their think-aloud processes and post-task reflections. 

\textit{All} participants ``jumped'' (made large movements across the map) in Artographer. We interpret jumping as divergent exploration--- motivated by participants' desire to collect a diversity of images, or a sense that they had not yet explored enough of the map. 
\begin{quote}
    ``I was trying to look for diversity of things
    ...kind of jumping around... these are all portraits of a similar variety, if I want something totally different, I just kind of zoom out and pan to a totally different region, or like, the Image Generator is almost like a roll of the dice, you know? It would help me to get out of a certain area, like, not stay in this region, go to a totally different--- portraits, landscapes, to still life.'' (P7)
\end{quote}
Here, P7 describes both manually jumping, and using image generation for ``targeted'' jumping --- e.g., prompting for ``a peaceful lake'' to generate an image that would, ideally, place them in an area around similarly peaceful landscape images --- without having to locate and navigate to that area manually.

We identified ``wandering'' behavior in 9/20 participants, characterizing this as a series of small movements across space, visiting nearby artworks along the way. We characterize this as more gradually controlled divergent exploration---as participants wandered, they described their attention being pulled towards nearby artworks that drew their eye. 
P12's movements in Figure \ref{fig:wandering} are exemplary.  Prolific wanderers (P12, P14, P8) generated relatively few images (0-1) and interacted with more artworks compared to other participants (see Figure \ref{fig:manual_vs_generating}), allowing curiosity and moments of piqued interest to direct the exploration. 
We identified ``revisiting'' behavior in 11/20 participants' free exploration trajectories---participants intentionally seeking and returning to an area they remembered previously visiting. 
We interpret this as a transition from divergent to convergent exploration: after broadly evaluating available areas, intentionally returning to one they've assessed as likely to yield more objects of interest to them. We consider how the affordances of visual-spatial interface navigation can uniquely support revisiting areas of interest in \S~\ref{sec:visual-spatial}.

We also identified ``fixation'' behavior, where participants spent a large portion of their time and collected many artworks in and around a single area. In Figure \ref{fig:designers}, we show the free exploration trajectories of 3 participants (P3, P8, P9) who all discussed a personal interest in abstract art. Paintings tagged ``abstract art'' and ``abstract painting'' are located in a central-westernmost region of the map, outlined in red. 
While all three of these fixated participants \textit{jumped} or \textit{wandered} away from the area, they all eventually \textit{revisited} it, with P8 \textit{only} collecting artworks within this region of interest. 
\subsection{Tensions and Opportunities Surfaced by Spatial Map Exploration} \label{sec:visual-spatial}
Participants described how exploring a computationally-constructed map of art space surfaced unique affordances and tensions.
\subsubsection{Visual-spatial navigation affords tacit media engagement.}
Participants used visual-spatial sense to tacitly navigate and explore media in Artographer---a well-reported affordance of 2D-spatial exploration interfaces. Participants described engaging with \textit{visual-spatial memory }and cues (e.g., region colors, images as ``landmarks'') to navigate effectively. As they quickly navigated back to one of their favorite images, P4 remarked on the novelty of the experience:
\begin{quote}
    ``It's like, I\textit{ know where I am}. Like, trying to find this Benjamin Franklin\footnote{Here, P4 was referring to \textit{Benjamin Franklin Drawing Electricity from the Sky} (1816) by Benjamin West}, it was in this sort of area, and there's a picture down here below of the Pylades\footnote{Here, P4 was referring to \textit{Pylades and Orestes Brought as Victims before Iphigenia} (1766), also by Benjamin West.}... there's this, like, visual memory that I don't think I've had in any other sort of artistic searching or medium before...
    that was kind of a trippy to have that jog my memory, just coming back to this area.'' (P4)
\end{quote}
P6 and P1 described feeling intrinsically motivated to develop their visual-spatial sense: 
 \begin{quote}
   ``If I had another half hour to spend with the system, I think I would... start to develop a sense of knowing the landscape, and being like, okay, I'm looking for this, so I know I need to go into the Eastern side of this specific map...it would feel really good to say that I have a grasp on how to navigate these, like, 15,000 artworks.'' (P6)
\end{quote}
\begin{quote}
    ``With more use, I would sort of pick up on the different affordances that I could use to keep my place in the map... like, associate the theme with the color that's going on there.'' (P1)
\end{quote}
\subsubsection{Exploring at the margins.}
\begin{figure*}
    \centering
    \includegraphics[width=0.87\textwidth]{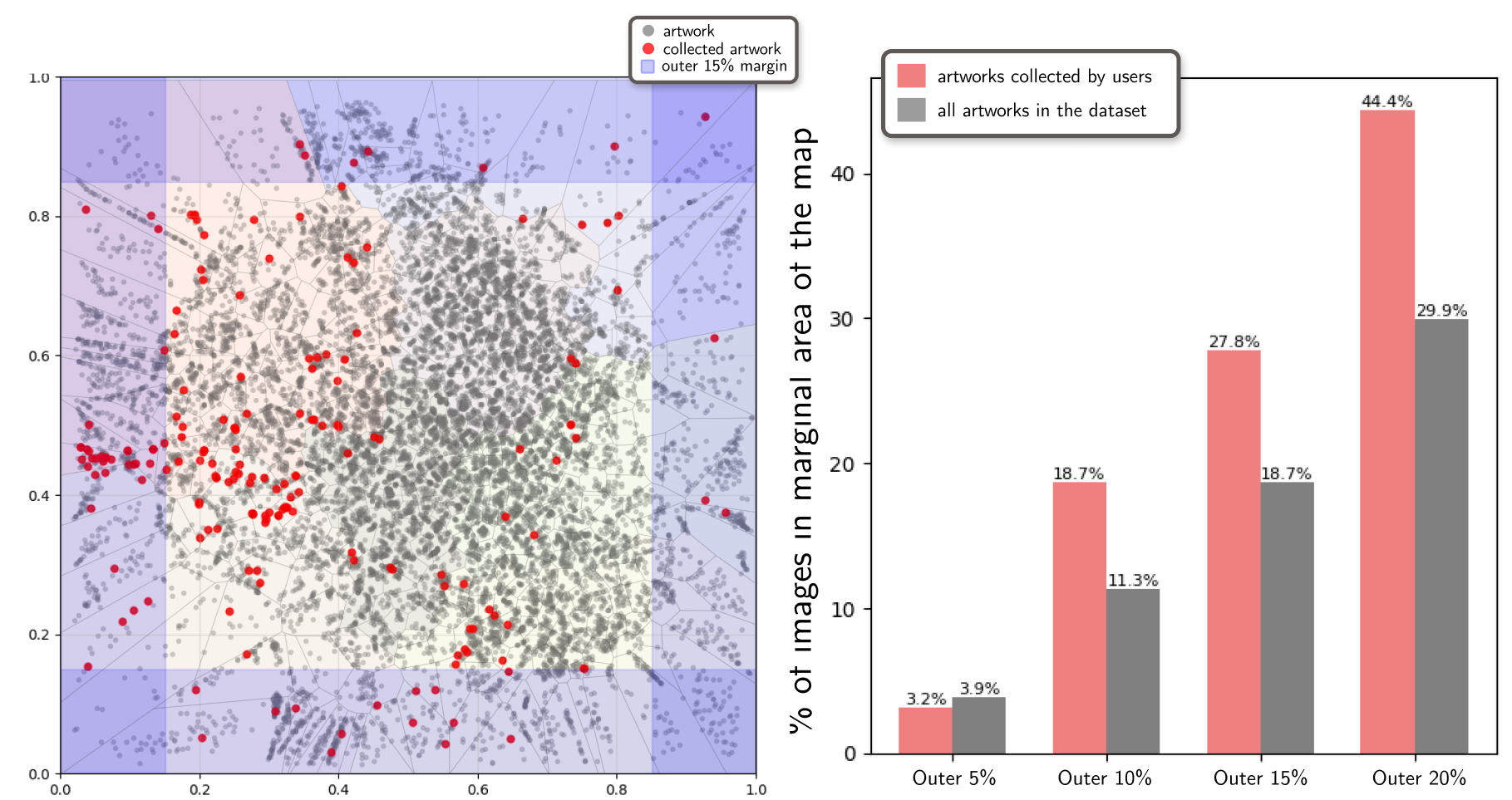}
    \caption{Participants collected a disproportionate number of artworks from the marginal edges of the spatial map.}
    \Description{This figure reveals a significant pattern in how participants explored the artwork collection, showing pronounced bias toward artworks positioned at the map's edges versus the center. The top visualization displays the spatial distribution of collected artworks. Gray dots represent all 15,958 artworks, while red dots indicate the 287 artworks collected by participants across all tasks. The map uses a purple-shaded background with a central light area transitioning to darker edges. Red dots cluster noticeably along the perimeter, particularly in the corners and edges, while the dense central region shows proportionally fewer collections despite containing the majority of the artworks. The bottom bar chart quantifies this edge bias by analyzing marginal areas of increasing size. The outer 5 percent of the map (furthest edges) shows near-equal representation—3.2 percent of collected artworks come from this region which contains 3.9 percent of all artworks. However, the bias intensifies dramatically in larger marginal areas. The outer 10 percent contains 11.3 percent of all artworks but accounts for 18.7 percent of collections. The outer 15 percent holds 18.7 percent of artworks but represents 27.8 percent of collections. Most strikingly, the outer 20 percent contains 29.9 percent of all artworks but accounts for 44.4 percent of collected items—nearly half of all selections despite representing less than a third of available options.}
    \label{fig:outliers}
\end{figure*}
Most (>80\%) of the artworks in the dataset are placed in the center 70\% of the spatial map--- yet participants collected a disproportionate number of artworks from its marginal edges (See Figure \ref{fig:outliers}). 
For some participants, this style of exploration was intentional: P4 described an interest in ‘the outer fringes' and others (P6, P10, P16, P19) intentionally avoided the map’s center. (The center is densely populated with Western portraiture, reflecting its over-representation in the WikiArt dataset, and in many historical art collections.)
\begin{quote}
``I'm trying to stray away from, like, faces. Once I get to the faces, which I've kind of hit now, I have to leave.'' (P10)
\end{quote}
P16, an art historian, aptly notes how data-driven spatial visualizations can reveal and reenact patterns of marginalization, and the dynamics that might arise:
\begin{quote}
    ``I'm very interested in the, like, center and periphery dynamic that's happening here... almost any collection you get is gonna put all your non-Western art around the edges, which means you get much less, like, vigorous and interesting connections between it and other things...'' (P16)
\end{quote} 
‘Fringe’ regions present a sparsity of relationships, and risk presenting richly diverse sets of artworks as if they are reductively similar. When AI models are used to automate art curation, these issues \textit{implicitly} impact viewers' experiences. Our participants' behavior shows that when this marginalia is made literally and explicitly accessible, users \textit{choose} to prod at these edges, and often find things of interest to them there. 
\label{sec:outliers}

\subsubsection{Participants engaged in critical conversation with the constructed presentation of art space.}
Multimodal embeddings allow us to surface visual and semantic relationships between artworks that are difficult to put into words. This does not stop people from trying---rather, the spatial map encouraged participants to critically interrogate how it was constructed. 
\begin{quote}
    ``One thing that I feel like I wasn't super sure about ...  \textit{how} it's actually placing the different regions, because some of them feel like they're based on, like....content, some of them are based around, like, the artist or the time period...'' (P6)
\end{quote}

\noindent Questioning the ``reasoning'' behind the map's structure was common across participants' thinking-aloud. 
They expressed \textit{agreeing} or \textit{disagreeing} with the system's placement ``choices'':
\begin{quote}
     ``I do agree with this characterization that this cat is more in the style of this cluster than this style. However, does this cat relate to Kandinsky at all? I'm going to say no.'' (P19)
\end{quote}
Navigating the map entails iteratively sensemaking its underlying structure; participants become attuned to moments of disorientation~\cite{biggs_thrown_2024}---when the map's construction did \textit{not} align with their expectations. These tensions were particularly noted by art history experts like P14, who each brought rich, preconstructed frameworks into the task of navigating a collection of historical artworks:
\begin{quote}
     ``The way the map is inherently spatialized... It brings together works by visual similarities that would otherwise not be considered `alike' in typical art historical methodologies.'' (P14)
\end{quote}
P14 took issue with the map's structure---then, upon reflection, began to challenge their expectations as \textit{constructed} by traditional art history contexts:
\begin{quote}
    ``This is just what I was taught... Art history people...would get really mad if you tried to talk about things just based on visual or thematic similarity, like, ``oh, this is ahistorical, that you're trying to make a comparison between these two things if they happened at different times in different places in the world... but because we're so focused on Western art history, there \textit{are} actually other connections between these different places, which may not geographically be next to each other...this can create new ways of engaging in art history, and that's super cool.'' (P14)
\end{quote}
\subsubsection{Participants noted novel tensions at the juxtaposition of historical and AI-generated imagery.}
\label{sec:aigenerated_juxtapositions}
Artographer presents AI image generation as a tool for augmenting navigation of the art space. It then enacts a uniquely dissonant media presentation, as AI-generated images are placed in context with historical artworks. Two art historians (P14, P16) discussed how this could be confusing or contribute to ongoing AI-driven media literacy and misinformation concerns. While we adjusted the presentation of AI generated images to visually distinguish them, participants \textit{did} confuse them:
\begin{quote} 
``I forgot I generated that one. It looks pretty cool, like, in relation to the others, though.'' (P18)
\end{quote}
Among the participants, P3 had limited experience studying art, and extensive experience and enthusiasm around AI image generation. During their free exploration, they prompted for an image of ``beautiful nature scenery''---the generated result placed them in an area of the map populated by landscape paintings.
\begin{quote}
    ``It almost looks like they're all Midjourney generated...
    I've seen a lot of AI-generated picture images that now it's hard to tell if these are AI-generated or not. But... oh, wow. I like these.
if you showed me this, I would say that this was AI-generated. 
I didn't know that this would be real.'' (P3)
\end{quote}
All of the ``favorite'' images that P3 submitted at the end of the study were discovered in this moment. They explained:
\begin{quote}
``It's interesting how people actually drew this in the past, and these aren't actually AI-generated, and it looks really, like it really requires talent.'' (P3)
\end{quote}
Here, P3’s interest in AI-generated landscapes served as an entrypoint to artworks they might not have otherwise discovered. Though these paintings are visually similar to images they encounter frequently, P3 found unique value in them because of their historical context as human-made artworks. 
We discuss in \S\ref{sec:serendipity} how future work might facilitate more serendipitous discoveries by helping users connect AI-generated imagery to rich and meaningful human contexts---while echoing art history experts' concerns that any presentation of AI-generated imagery should be intentional in its communication and treatment of image provenance. 

%% file: sections/5-discussion.tex
\section{Curatorial Values for Media Interface Design}
We used Artographer as a probe, inviting community members to help us investigate how system design can shape media reception and engagement experiences. Reflecting on our findings, we identify four \textit{curatorial} design values that surface across our participants engagements with artwork in this system: visibility, agency, friction, and serendipity. We present and apply these values as a critical lens on the design of interfaces that select and present creative work, towards challenging dominant media platform design paradigms. 
\begin{table*}[ht]
\centering
\scriptsize
\setlength{\tabcolsep}{8pt}
\renewcommand{\arraystretch}{1.5}
\begin{tabular}{|p{2cm}|p{3.5cm}|p{3.5cm}|p{3.5cm}|}
\hline
\textbf{Curatorial Design Values} & \textbf{Similarity-Clustered Spatial Map} & \textbf{Query-Based Search} & \textbf{Recommendation Feed} \\
\hline
\textbf{1. Visibility}

How well can the user perceive the space of options available? & 
\textbf{More:} Ideally, the user can quickly gauge what is generally available, i.e., at a distant read. & 
\textbf{Less:} Ideally, the user rarely sees options outside the specific space that they explicitly expressed interest in. & 
\textbf{Less:} Ideally, the user rarely sees content outside the space of options algorithmically determined as interesting to them. \\
\hline
\textbf{2. Agency}

Does the user have control over \textit{where} and \textit{how} to look? & 
\textbf{More:} Users can, and typically \textit{must}, directly "move" themselves to the place they want to look next & 
\textbf{More:} Ideally, the user is able to precisely specify how and where to look for results. & 
\textbf{Less:} Ideally, the recommendation algorithm takes on the labor of seeking out content on behalf of the user. \\

\hline
\textbf{3. Friction}

How much work is required for the user to get to something they are interested in? & 
\textbf{More:} The user must take an active, reflective role in searching for pieces that interest them. & 
\textbf{Less:} Ideally, the user can easily and precisely specify what they are looking for, and find it among the top results. & 
\textbf{Less:} Ideally, the user experiences minimal friction---they may not even perceive a "search" is happening. \\

\hline
\textbf{4. Serendipity}

Can the user find interesting things that they were not intentionally searching for? & 
\textbf{More:} Visibility draws users away from intentions, towards areas of interest; ideally, clustering ensures that other interesting results are frequently visible nearby. & 
\textbf{Less:} Ideally, the system presents results that are directly related to the user's intentionally specified search query. & 
\textbf{More:} Ideally, the user only needs to express minimal or even incidental intention, for the system to then continually present them with interesting results. \\

\hline
\textbf{Designs for...} & Exploration, active discovery & Targeted search & Passive consumption \\
\hline
\end{tabular}
\caption{This table demonstrates application of the four curatorial design values we identified--- Visibility, Agency, Friction, and Serendipity--- as an evaluative lens on three interfaces that select and present media.}
\label{tab:discovery_comparison}
\Description{This table presents a comparative analysis of three curatorial interface types—similarity-clustered spatial maps, query-based search, and recommendation feeds—evaluated across four key design values that shape user exploration experiences. The table uses text labels of "More" and "Less" to indicate where each interface type excels or has limitations across each design dimension. For Visibility, spatial maps excel by allowing users to perceive the full scope of available options at a distant read, providing immediate awareness of the collection's breadth. Query-based search limits visibility to explicitly requested results, while recommendation feeds restrict users to algorithmically selected content, potentially creating filter bubbles. Regarding Agency, spatial maps require users to actively navigate and directly control their exploration path through deliberate movement decisions. Query-based search offers high agency, enabling precise specification of search parameters. Recommendation feeds minimize agency, as algorithms determine content presentation based on inferred user preferences rather than explicit choices. For Serendipity, spatial maps facilitate unexpected discoveries through their visibility and clustering properties—interesting but unintended content naturally appears adjacent to targeted items. Query-based search minimizes serendipity by returning only directly relevant results. Recommendation feeds paradoxically achieve high serendipity through algorithmic suggestion, requiring minimal user intention to surface diverse content. Concerning Friction, spatial maps demand the most user effort, requiring active, reflective searching through navigation and visual scanning. Query-based search reduces friction by enabling direct specification of needs with immediate results. Recommendation feeds minimize friction entirely, creating a passive consumption experience where content appears without conscious searching. The bottom row synthesizes these characteristics: spatial maps best support exploration and active discovery, query-based search optimizes targeted information retrieval, and recommendation feeds enable passive content consumption.}
\end{table*}

\subsection{Visibility}
\begin{quote}
     ``Maybe I’ll zoom out to see what else is available to me [...] It's really cool for just exploration, just seeing all the different artworks that are out there.'' (P18)
\end{quote}
A spatial map interface provides unique affordances for participants to view what is \textit{available}. 
Visibility allowed users to navigate to and from areas of interest, and make choices informed by the map's higher-level structure (e.g., choosing to explore its edges) and to \textit{interrogate} its structures.

Systems that make creative activities more visible and traceable hold \textit{methodological} value for creativity research. Prior work has proposed the use of \textit{expressive range analysis} and other creative activity tracing (CAT) techniques to visualize and evaluate how a tool supports access and exploration of a possibility space~\cite{smith_analyzing_2010, kreminski_evaluating_2022, hammad_tracing_2026}. 
Visible and tangible evaluation instruments can give researchers \textit{and participants} real-time insights into participants creative behavior~\cite{isbister_sensual_2006, hammad_towards_2025}. Developing systems as substrates for participatory creative experimentation can help researchers capture insights that go beyond quantification-based evaluation instruments (e.g,. the CSI~\cite{cherry_quantifying_2014}), towards richer, more comprehensive understanding of creative activity. 
\subsection{Agency}
\label{sec:agency}

Prior work has identified how dominant social platforms can deteriorate users' sense of \textit{agency} in media interactions~\cite{lukoff_how_2021, baughan_i_2022, baumer_departing_2018}. Scrolling content feeds curated by `black box'' AI recommendation systems effectively \textit{obscure} both the space of media available on the platform, and the underlying structures relating them. A recommendation is, implicitly, a ``nearest neighbor'' of some other option(s) the user expressed interest in. However, users are rarely given insight into the underlying assumptions constructing this space of recommendations, let alone the opportunity to act on them. 
Recommendation systems also tend to funnel users towards ``rabbitholes'' or narrow areas of interest. 

Participants' reflective engagements with media in Artographer illuminated how being empowered to \textit{choose} where, and how, to explore next is meaningful. 
Participants who intentionally chose to fixate on a group of artworks also \textit{chose} to \textit{jump} to other areas of art space, before \textit{revisiting} that area of interest (as in Figure \ref{fig:designers}.) 
Designing for visibility and agency in media engagement empowers users to participate in the informed \textit{decision} to stay in a narrow area of interest---the option to explore elsewhere is meaningful, even when rejected. 
\subsection{Friction}
\label{sec:friction}
\edits{Navigating the Artographer map required users to continuously evaluate the artworks it presented to them, and to actively steer their explorations; artworks sometimes clustered and occluded one another, and users sometimes got ``lost'' or distracted.} Artographer's interface enacted \textbf{friction} and \textit{seamfulness}. As a move against long-prevailing  ``seamless'' interaction design standards, these qualities show potential for encouraging \textit{active} (rather than passive) user engagement~\cite{hook_strong_2012, sheahan_designing_2024}, and for empowering more critical engagement with AI systems~\cite{liu_agency_2025, ehsan_seamful_2024}. 

Intentionally designing for friction and \textit{seams} in art-finding can help facilitate moments where users can slow down, evaluate, interrogate and relate to the work being presented. 
\edits{Seamful design of media platforms could also explore \textit{exposing} the limits and inherent biases of the curation systems that underly them, inviting users to perceive, and interpretively critique~\cite{bardzell_interaction_2008} their constructedness.}
\edits{Beyond designing media platforms for \textit{ease} of art discovery and consumption---designing for frictive artistic encounters could help facilitate the kind of ``engagement'' necessary for a more critical, thoughtful, and reflective~\cite{kreminski_reflective_2021} media ecosystem to thrive.}

\subsection{Serendipity in the Art of Art-Finding}
\label{sec:serendipity}
Where HCI CST research has often emphasized the production of artifacts~\cite{compositiontools, rhys_cox_beyond_2025, li_beyond_2023}, 
Artographer emphasizes process---centering creative exploration and engagement with artwork as an intrinsically valuable~\cite{CasualCreators}, reflective~\cite{kreminski_reflective_2021, glinka_critical-reflective_2023}, and dialectical~\cite{zhang_searching_2024} activity. 
How can we support the autotelic art-finding process---designing distribution platforms that can facilitate more personally fulfilling, meaningful encounters with media? In shifting away from centering platform values, we might move toward centering \textbf{serendipity. }

Participants found it \textit{fun} to freely explore in the spatial map interface. They connected this to the feeling of serendipity---delighting in the discovery of something they did not expect to find or learn about. 
7/20 participants described surprise or serendipity as the reason why they selected one or more of their favorite images at the end of the free exploration task. 

We suggest that Serendipity is \textit{amplified} by Visibility, Friction, and Agency: when participants exert active effort into navigating a space of artworks, they earn a sense of ownership over their path of navigation and any discoveries they find along the way. 

Serendipity has appeared as a value in other engaging curatorial  interfaces \cite{meyer_algorithmic_2024, frost_art_2019} \textit{and} in search-based recommendation systems~\cite{ninomiya_determinants_2025}.  
In systems that maintain media \textit{provenance}~\cite{almeda_creativity_2025}, serendipitous discoveries can seed personal connection and appreciation for specific aesthetics, artworks, and \textit{human artists.}
\begin{quote}
    ``I found, like, 5 artists today I didn't know about that I want to research more.'' (P4)
\end{quote}
P19 describes a new, joyful appreciation for Louis Wain:
\begin{quote}
    ``...it just had more surprises. I feel like the [map] interface is a really great learning tool... even for me, who I feel like I'm somewhat knowledgeable about art.... it was such a joy for me to learn about these artists, to discover new artists like this person who only drew cats in the mental hospital...'' (P19)
\end{quote}
\noindent
Several participants in our study exhibited behaviors in Artographer that \citeauthor{nelson_curious_2018} describe as characteristic of ``curious users of casual creators,'' e.g., making larger leaps in the design space or prodding at its edges~\cite{nelson_curious_2018}. \textit{Curious use} is intrinsically motivated by curiosity in the space of possibilities available for exploration, in contrast with goal-driven or artifact-driven use. Future work might explore how aspects of interface design can best pique curiosity---scaffolding the goals of curious users, or facilitating serendipity by drawing users toward underexplored areas.

P3's experience discovering an unexpected appreciation for historical landscape paintings (see  \ref{sec:aigenerated_juxtapositions}) points to opportunities for future work on AI systems to help users draw connections from the media and aesthetics they are interested in, to established sources of potential cultural or historical relevance. Designing for this kind of \textit{fuzzy provenance }and serendipity could allow AI-CSTs to contribute to sustaining human-artmaking and appreciation in creative ecosystems, rather than disrupting and \textit{extracting} from them~\cite{almeda_creativity_2025}.

%% file: sections/7-conclusion.tex
\section{Conclusion}
\begin{quote}
    \textit{
    What happens when a new work of art is created is something that happens simultaneously to all the works of art which preceded it.
    The existing monuments form an ideal order among themselves, which is modified by the introduction of the new (the really new) work of art among them...
    the past should be altered by the present as much as the present is directed by the past.} \begin{flushright}---T. S. Eliot\end{flushright}
\end{quote}
    
\noindent
In this work, we presented Artographer: a design exploration into the curation and presentation of media. Artographer presents an intentionally curated dataset of 15,958 historical artworks as a zoomable, similarity-clustered spatial map, alongside a contextualized text-to-image generation system that places AI-generated images in conversation with ``nearby'' historical artworks. We instrumented Artographer as a probe in an empirical study, using it to capture and trace art exploration activity, and to elicit perspectives on how presentation system design can shape artwork reception and engagement. We characterize how 20 participants, including 9 art history scholars, jumped, wandered, became fixated, and revisited familiar areas across this particular space of artworks. We discuss four curatorial values: Visibility, Agency, Friction, and Serendipity---and consider what we lose when media systems automate away from these values.
We advocate for treating the rich, relational knowledge embedded into AI models not as a ``solution'' for human subjectivity in curation, but as a flexible material that users might joyfully manipulate, and reflexively interrogate. We draw from hypertext artist Chia Amisola's framing of media navigation as a form of creative authorship itself~\cite{amisola_becoming_2024}, and consider how the design of alternative distribution systems can help us move from a media ecosystem dominated by content \textit{recommendation} systems towards artistic \textit{curatorial} systems: reimagining a future of digital media platforms that treat the creative work they distribute with care, and that empower users with the agency to navigate more meaningful, critical engagements with media. 

%% file: sections/manual_references.bib
@article{cherry_quantifying_2014,
	title = {Quantifying the {Creativity} {Support} of {Digital} {Tools} through the {Creativity} {Support} {Index}},
	volume = {21},
	issn = {1073-0516},
	url = {https://dl.acm.org/doi/10.1145/2617588},
	doi = {10.1145/2617588},
	number = {4},
	urldate = {2025-09-12},
	journal = {ACM Trans. Comput.-Hum. Interact.},
	author = {Cherry, Erin and Latulipe, Celine},
	month = jun,
	year = {2014},
	pages = {21:1--21:25},
}

@misc{valyaeva_ai_2023,
	title = {{AI} {Image} {Statistics} for 2024: {How} {Much} {Content} {Was} {Created} by {AI}},
	shorttitle = {{AI} {Image} {Statistics} for 2024},
	url = {https://journal.everypixel.com/ai-image-statistics},
	language = {en-US},
	urldate = {2025-09-06},
	journal = {Everypixel Journal},
	author = {Valyaeva, Alina},
	month = aug,
	year = {2023},
}

@misc{artsy_artsy-art-genome-project_2025,
	address = {artsy/the-art-genome-project: Gene names and definitions},
	title = {Artsy/{The}-{Art}-{Genome}-{Project}},
	url = {https://github.com/artsy/the-art-genome-project/tree/master},
	urldate = {2025-09-11},
	publisher = {GitHub},
	author = {Artsy},
	year = {2025},
}

@article{an_art_2024,
	title = {Art curation in virtual spaces: {The} influence of digital technology in redefining the aesthetics and interpretation of art {\textbar} {Humanities}, {Arts} and {Social} {Sciences} {Studies}},
	url = {https://so02.tci-thaijo.org/index.php/hasss/article/view/267552},
	doi = {https://doi.org/10.69598/hasss.24.2.267552},
	urldate = {2025-09-11},
	journal = {Humanities, Arts and Social Sciences Studies},
	author = {An, Ran},
	month = aug,
	year = {2024},
}

@inproceedings{von_davier_designing_2023,
	address = {New York, NY, USA},
	series = {{CHI} {EA} '23},
	title = {Designing for {Appreciation}: {How} {Digital} {Spaces} {Can} {Support} {Art} and {Culture}},
	isbn = {978-1-4503-9422-2},
	shorttitle = {Designing for {Appreciation}},
	url = {https://dl.acm.org/doi/10.1145/3544549.3577041},
	doi = {10.1145/3544549.3577041},
	urldate = {2025-09-10},
	booktitle = {Extended {Abstracts} of the 2023 {CHI} {Conference} on {Human} {Factors} in {Computing} {Systems}},
	publisher = {Association for Computing Machinery},
	author = {Von Davier, Thomas Serban},
	month = apr,
	year = {2023},
	pages = {1--5},
}

@misc{jeon2025dimensionality,
      title={Dimensionality Reduction Considered Harmful (Some of the Time)}, 
      author={Hyeon Jeon},
      year={2025},
      eprint={2512.18230},
      archivePrefix={arXiv},
      primaryClass={cs.HC},
      url={https://arxiv.org/abs/2512.18230}, 
}

@misc{coenen_understanding_2025,
	title = {Understanding {UMAP}},
	url = {https://pair-code.github.io/understanding-umap/},
	urldate = {2025-09-11},
	journal = {Understanding UMAP},
	author = {Coenen, Andy and Pearce, Adam},
	year = {2025},
}

@article{villaespesa_museum_2019,
	title = {Museum {Collections} and {Online} {Users}: {Development} of a {Segmentation} {Model} for the {Metropolitan} {Museum} of {Art}},
	volume = {22},
	issn = {1064-5578},
	shorttitle = {Museum {Collections} and {Online} {Users}},
	url = {https://doi.org/10.1080/10645578.2019.1668679},
	doi = {10.1080/10645578.2019.1668679},
	number = {2},
	urldate = {2025-09-10},
	journal = {Visitor Studies},
	author = {Villaespesa, Elena},
	month = jul,
	year = {2019},
	note = {Publisher: Routledge
\_eprint: https://doi.org/10.1080/10645578.2019.1668679},
	pages = {233--252},
}

@inproceedings{meyer_algorithmic_2024,
	address = {Honolulu HI USA},
	title = {Algorithmic {Ways} of {Seeing}: {Using} {Object} {Detection} to {Facilitate} {Art} {Exploration}},
	isbn = {979-8-4007-0330-0},
	shorttitle = {Algorithmic {Ways} of {Seeing}},
	url = {https://dl.acm.org/doi/10.1145/3613904.3642157},
	doi = {10.1145/3613904.3642157},
	language = {en},
	urldate = {2025-09-10},
	booktitle = {Proceedings of the {CHI} {Conference} on {Human} {Factors} in {Computing} {Systems}},
	publisher = {ACM},
	author = {Meyer, Louie and Aaen, Johanne Engel and Tranberg, Anitamalina Regitse and Kun, Peter and Freiberger, Matthias and Risi, Sebastian and Løvlie, Anders Sundnes},
	month = may,
	year = {2024},
	pages = {1--18},
}

@inproceedings{kreminski_evaluating_2022,
	address = {Helsinki, FInland},
	title = {Evaluating {Mixed}-{Initiative} {Creative} {Interfaces} via {Expressive} {Range} {Coverage} {Analysis}},
	url = {https://mkremins.github.io/publications/ERaCA_HAI-GEN2022.pdf},
	language = {en},
	urldate = {2025-09-04},
	booktitle = {Joint {Proceedings} of the {ACM} {IUI} {Workshops} 2022},
	publisher = {ACM},
	author = {Kreminski, Max and Karth, Isaac and Mateas, Michael and Wardrip-Fruin, Noah},
	month = mar,
	year = {2022},
}

@inproceedings{smith_analyzing_2010,
	address = {New York, NY, USA},
	series = {{PCGames} '10},
	title = {Analyzing the expressive range of a level generator},
	isbn = {978-1-4503-0023-0},
	url = {https://dl.acm.org/doi/10.1145/1814256.1814260},
	doi = {10.1145/1814256.1814260},
	urldate = {2025-09-10},
	booktitle = {Proceedings of the 2010 {Workshop} on {Procedural} {Content} {Generation} in {Games}},
	publisher = {Association for Computing Machinery},
	author = {Smith, Gillian and Whitehead, Jim},
	month = jun,
	year = {2010},
	pages = {1--7},
}

@inproceedings{frost_art_2019,
	address = {Boston, MA},
	title = {Art {I} {Don}’t {Like}: {An} {Anti}-{Recommender} {System} for {Visual} {Art}},
	shorttitle = {Art {I} {Don}’t {Like}},
	url = {https://mw19.mwconf.org/proposal/art-i-dont-like-an-anti-recommender-system-for-visual-art/index.html},
	language = {en-US},
	urldate = {2025-09-09},
	booktitle = {{MW19}, the 23rd annual {MuseWeb} conference},
	author = {Frost, Sarah and Thomas, Manu Mathew and Forbes, Angus G.},
	month = apr,
	year = {2019},
}

@misc{schuhmann_laion-aesthetics_2022,
	title = {{LAION}-{Aesthetics}},
	url = {https://laion.ai/blog/laion-aesthetics},
	language = {en},
	urldate = {2025-09-07},
	journal = {LAION},
	author = {Schuhmann, Christoph},
	month = aug,
	year = {2022},
}

@misc{beaumont_laion-aiaesthetic-predictor_2022,
	title = {{LAION}-{AI}/aesthetic-predictor},
	copyright = {MIT},
	url = {https://github.com/LAION-AI/aesthetic-predictor},
	urldate = {2025-09-07},
	publisher = {LAION AI},
	author = {Beaumont, Romain and Schuhman, Christoph},
	month = may,
	year = {2022},
	note = {original-date: 2022-05-21T12:36:24Z},
}

@misc{amisola_becoming_2024,
	title = {Becoming hypertext},
	url = {https://everythingi.love},
	language = {en},
	urldate = {2025-09-10},
	journal = {Everything I Love {\textbar} Chia Amisola},
	author = {Amisola, Chia},
	month = dec,
	year = {2024},
}

@article{baumer_departing_2018,
	title = {Departing and {Returning}: {Sense} of {Agency} as an {Organizing} {Concept} for {Understanding} {Social} {Media} {Non}/use {Transitions}},
	volume = {2},
	shorttitle = {Departing and {Returning}},
	url = {https://dl.acm.org/doi/10.1145/3274292},
	doi = {10.1145/3274292},
	number = {CSCW},
	urldate = {2025-09-10},
	journal = {Proc. ACM Hum.-Comput. Interact.},
	author = {Baumer, Eric P. S. and Sun, Rui and Schaedler, Peter},
	month = nov,
	year = {2018},
	pages = {23:1--23:19},
}

@inproceedings{lukoff_how_2021,
	address = {New York, NY, USA},
	series = {{CHI} '21},
	title = {How the {Design} of {YouTube} {Influences} {User} {Sense} of {Agency}},
	isbn = {978-1-4503-8096-6},
	url = {https://dl.acm.org/doi/10.1145/3411764.3445467},
	doi = {10.1145/3411764.3445467},
	urldate = {2025-09-09},
	booktitle = {Proceedings of the 2021 {CHI} {Conference} on {Human} {Factors} in {Computing} {Systems}},
	publisher = {Association for Computing Machinery},
	author = {Lukoff, Kai and Lyngs, Ulrik and Zade, Himanshu and Liao, J. Vera and Choi, James and Fan, Kaiyue and Munson, Sean A. and Hiniker, Alexis},
	month = may,
	year = {2021},
	pages = {1--17},
}

@inproceedings{ohm_collection_2023,
	address = {New York, NY, USA},
	series = {{VINCI} '23},
	title = {Collection {Space} {Navigator}: {An} {Interactive} {Visualization} {Interface} for {Multidimensional} {Datasets}},
	isbn = {979-8-4007-0751-3},
	shorttitle = {Collection {Space} {Navigator}},
	url = {https://dl.acm.org/doi/10.1145/3615522.3615546},
	doi = {10.1145/3615522.3615546},
	urldate = {2025-09-10},
	booktitle = {Proceedings of the 16th {International} {Symposium} on {Visual} {Information} {Communication} and {Interaction}},
	publisher = {Association for Computing Machinery},
	author = {Ohm, Tillmann and Canet Sola, Mar and Karjus, Andres and Schich, Maximilian},
	month = oct,
	year = {2023},
	pages = {1--5},
}

@inproceedings{kreminski_herding_2026,
	address = {New York, NY, USA},
	series = {{CHI} {EA} '26},
	title = {Herding {CATs}: {Making} {Sense} of {Creative} {Activity} {Traces}},
	isbn = {979-8-4007-2281-3},
	shorttitle = {Herding {CATs}},
	url = {https://dl.acm.org/doi/10.1145/3772363.3778743},
	doi = {10.1145/3772363.3778743},
	urldate = {2026-05-05},
	booktitle = {Proceedings of the {Extended} {Abstracts} of the 2026 {CHI} {Conference} on {Human} {Factors} in {Computing} {Systems}},
	publisher = {Association for Computing Machinery},
	author = {Kreminski, Max and Smith, Amy and Joon Young Chung, John and Son, Kihoon and Yang, Qian and Lee, Sang Won and Hammad, Noor and Rawn, Eric and Garanganao Almeda, Shm and Zamfirescu-Pereira, J.D.},
	month = apr,
	year = {2026},
	pages = {1--5},
}

@inproceedings{hammad_tracing_2026,
	address = {New York, NY, USA},
	series = {{CHI} '26},
	title = {Tracing {Creativity}: {A} {Design} {Space} {For} {Creative} {Activity} {Traces} in {HCI}},
	isbn = {979-8-4007-2278-3},
	shorttitle = {Tracing {Creativity}},
	url = {https://dl.acm.org/doi/10.1145/3772318.3791263},
	doi = {10.1145/3772318.3791263},
	urldate = {2026-05-05},
	booktitle = {Proceedings of the 2026 {CHI} {Conference} on {Human} {Factors} in {Computing} {Systems}},
	publisher = {Association for Computing Machinery},
	author = {Hammad, Noor and Lin, David Chuan-En and Smith, Amy and Kreminski, Max and Harpstead, Erik and Hammer, Jessica},
	month = apr,
	year = {2026},
	pages = {1--24},
}

@inproceedings{hammad_towards_2025,
	address = {New York, NY, USA},
	series = {{CHI} {PLAY} {Companion} '25},
	title = {Towards {Trail}-{Aware} {Digital} {Content} {Creation} {Tools}},
	isbn = {979-8-4007-2023-9},
	url = {https://dl.acm.org/doi/10.1145/3744736.3749328},
	doi = {10.1145/3744736.3749328},
	urldate = {2026-05-05},
	booktitle = {Companion {Proceedings} of the {Annual} {Symposium} on {Computer}-{Human} {Interaction} in {Play}},
	publisher = {Association for Computing Machinery},
	author = {Hammad, Noor},
	month = oct,
	year = {2025},
	pages = {273--275},
}

@article{hook_strong_2012,
	title = {Strong concepts: {Intermediate}-level knowledge in interaction design research},
	volume = {19},
	issn = {1073-0516},
	shorttitle = {Strong concepts},
	url = {https://dl.acm.org/doi/10.1145/2362364.2362371},
	doi = {10.1145/2362364.2362371},
	number = {3},
	urldate = {2026-04-01},
	journal = {ACM Trans. Comput.-Hum. Interact.},
	author = {Höök, Kristina and Löwgren, Jonas},
	month = oct,
	year = {2012},
	pages = {23:1--23:18},
}

@misc{oygard_visualizing_2018,
	title = {Visualizing an art collection},
	url = {http://auduno.github.io/2018/10/27/visualizing-an-art-collection/index.html},
	language = {en},
	urldate = {2026-05-06},
	journal = {Audun M. Øygard},
	author = {Øygard, Audun M.},
	month = oct,
	year = {2018},
}

@misc{pietsch_cpietschvikus-viewer_2026,
	title = {cpietsch/vikus-viewer},
	url = {https://github.com/cpietsch/vikus-viewer},
	urldate = {2026-05-06},
	author = {Pietsch, Christopher},
	month = apr,
	year = {2026},
	note = {original-date: 2018-05-12T17:40:59Z},
}

@misc{leonard_pleonard212pix-plot_2026,
	title = {pleonard212/pix-plot},
	copyright = {MIT},
	url = {https://github.com/pleonard212/pix-plot},
	urldate = {2026-05-06},
	author = {Leonard, Peter},
	month = apr,
	year = {2026},
	note = {original-date: 2017-08-03T18:25:19Z},
}

@article{colwell2023distant,
  title={Distant Viewing: Computational Image Similarity and Visual Resources Collections},
  author={Colwell, Tess and King, Lindsay},
  journal={Art Documentation: Journal of the Art Libraries Society of North America},
  volume={42},
  number={2},
  pages={182--194},
  year={2023},
  publisher={The University of Chicago Press Chicago, IL}
}

@misc{diagne_t-sne_2018,
	title = {t-{SNE} {Map} by {Cyril} {Diagne}, {Nicolas} {Barradeau} \&amp; {Simon} {Doury} - {Experiments} with {Google}},
	url = {https://experiments.withgoogle.com/t-sne-map},
	urldate = {2026-05-06},
	journal = {Experiments with Google},
	author = {Diagne, Cyril and {Simon Doury} and {Nicolas Barradeau}},
	month = mar,
	year = {2018},
}

@inproceedings{kaplan2016visual,
  title={Visual patterns discovery in large databases of paintings},
  author={Kaplan, Fr{\'e}d{\'e}ric},
  booktitle={Digital Humanities 2016},
  year={2016}
}

@inproceedings{boehner_how_2007,
	address = {New York, NY, USA},
	series = {{CHI} '07},
	title = {How {HCI} interprets the probes},
	isbn = {978-1-59593-593-9},
	url = {https://dl.acm.org/doi/10.1145/1240624.1240789},
	doi = {10.1145/1240624.1240789},
	urldate = {2026-05-05},
	booktitle = {Proceedings of the {SIGCHI} {Conference} on {Human} {Factors} in {Computing} {Systems}},
	publisher = {Association for Computing Machinery},
	author = {Boehner, Kirsten and Vertesi, Janet and Sengers, Phoebe and Dourish, Paul},
	month = apr,
	year = {2007},
	pages = {1077--1086},
}

@article{von_davier_looking_2025,
	title = {Looking for {Art} in a {Sea} of {Content}: {A} {Human}-{Centered} {Approach} to {Supporting} {Creativity} on {Social} {Media}},
	volume = {9},
	shorttitle = {Looking for {Art} in a {Sea} of {Content}},
	url = {https://dl.acm.org/doi/10.1145/3711025},
	doi = {10.1145/3711025},
	number = {2},
	urldate = {2025-09-09},
	journal = {Proc. ACM Hum.-Comput. Interact.},
	author = {von Davier, Thomas Serban and Noh, Hayoun and Van Kleek, Max and Shadbolt, Nigel},
	month = may,
	year = {2025},
	pages = {CSCW127:1--CSCW127:25},
}

@book{charmaz_constructing_2006,
	title = {Constructing {Grounded} {Theory}: {A} {Practical} {Guide} through {Qualitative} {Analysis}},
	isbn = {978-1-4462-0040-7},
	shorttitle = {Constructing {Grounded} {Theory}},
	language = {en},
	publisher = {SAGE},
	author = {Charmaz, Kathy},
	month = jan,
	year = {2006},
	note = {Google-Books-ID: 2ThdBAAAQBAJ},
}

@inproceedings{feng_mapping_2024,
	address = {New York, NY, USA},
	series = {{CHI} '24},
	title = {Mapping the {Design} {Space} of {Teachable} {Social} {Media} {Feed} {Experiences}},
	isbn = {979-8-4007-0330-0},
	url = {https://dl.acm.org/doi/10.1145/3613904.3642120},
	doi = {10.1145/3613904.3642120},
	urldate = {2026-02-05},
	booktitle = {Proceedings of the 2024 {CHI} {Conference} on {Human} {Factors} in {Computing} {Systems}},
	publisher = {Association for Computing Machinery},
	author = {Feng, K. J. Kevin and Koo, Xander and Tan, Lawrence and Bruckman, Amy and McDonald, David W. and Zhang, Amy X.},
	month = may,
	year = {2024},
	pages = {1--20},
}

@inproceedings{baughan_i_2022,
	address = {New York, NY, USA},
	series = {{CHI} '22},
	title = {“{I} {Don}’t {Even} {Remember} {What} {I} {Read}”: {How} {Design} {Influences} {Dissociation} on {Social} {Media}},
	isbn = {978-1-4503-9157-3},
	shorttitle = {“{I} {Don}’t {Even} {Remember} {What} {I} {Read}”},
	url = {https://dl.acm.org/doi/10.1145/3491102.3501899},
	doi = {10.1145/3491102.3501899},
	urldate = {2026-02-05},
	booktitle = {Proceedings of the 2022 {CHI} {Conference} on {Human} {Factors} in {Computing} {Systems}},
	publisher = {Association for Computing Machinery},
	author = {Baughan, Amanda and Zhang, Mingrui Ray and Rao, Raveena and Lukoff, Kai and Schaadhardt, Anastasia and Butler, Lisa D. and Hiniker, Alexis},
	month = apr,
	year = {2022},
	pages = {1--13},
}

@inproceedings{zhang_searching_2024,
	address = {New York, NY, USA},
	series = {{CHI} '24},
	title = {Searching for the {Non}-{Consequential}: {Dialectical} {Activities} in {HCI} and the {Limits} of {Computers}},
	isbn = {979-8-4007-0330-0},
	shorttitle = {Searching for the {Non}-{Consequential}},
	url = {https://dl.acm.org/doi/10.1145/3613904.3641945},
	doi = {10.1145/3613904.3641945},
	urldate = {2026-01-17},
	booktitle = {Proceedings of the 2024 {CHI} {Conference} on {Human} {Factors} in {Computing} {Systems}},
	publisher = {Association for Computing Machinery},
	author = {Zhang, Haoqi},
	month = may,
	year = {2024},
	pages = {1--13},
}

@article{dixon2025most,
  title={Most popular social networks worldwide as of February 2025, by number of monthly active users},
  author={Dixon, Stacey J},
  journal={Statista. March},
  volume={26},
  year={2025}
}

@inproceedings{zou_reinforcement_2019,
	address = {New York, NY, USA},
	series = {{KDD} '19},
	title = {Reinforcement {Learning} to {Optimize} {Long}-term {User} {Engagement} in {Recommender} {Systems}},
	isbn = {978-1-4503-6201-6},
	url = {https://dl.acm.org/doi/10.1145/3292500.3330668},
	doi = {10.1145/3292500.3330668},
	urldate = {2026-02-06},
	booktitle = {Proceedings of the 25th {ACM} {SIGKDD} {International} {Conference} on {Knowledge} {Discovery} \& {Data} {Mining}},
	publisher = {Association for Computing Machinery},
	author = {Zou, Lixin and Xia, Long and Ding, Zhuoye and Song, Jiaxing and Liu, Weidong and Yin, Dawei},
	month = jul,
	year = {2019},
	pages = {2810--2818},
}

@inproceedings{biggs_thrown_2024,
	address = {New York, NY, USA},
	series = {{CHI} '24},
	title = {Thrown from {Normative} {Ground}: {Exploring} the {Potential} of {Disorientation} as a {Critical} {Methodological} {Strategy} in {HCI}},
	isbn = {979-8-4007-0330-0},
	shorttitle = {Thrown from {Normative} {Ground}},
	url = {https://dl.acm.org/doi/10.1145/3613904.3642724},
	doi = {10.1145/3613904.3642724},
	urldate = {2026-01-10},
	booktitle = {Proceedings of the 2024 {CHI} {Conference} on {Human} {Factors} in {Computing} {Systems}},
	publisher = {Association for Computing Machinery},
	author = {Biggs, Heidi and Bardzell, Shaowen},
	month = may,
	year = {2024},
	pages = {1--11},
}

@article{ehsan_seamful_2024,
	title = {Seamful {XAI}: {Operationalizing} {Seamful} {Design} in {Explainable} {AI}},
	volume = {8},
	shorttitle = {Seamful {XAI}},
	url = {https://dl.acm.org/doi/10.1145/3637396},
	doi = {10.1145/3637396},
	number = {CSCW1},
	urldate = {2026-02-06},
	journal = {Proc. ACM Hum.-Comput. Interact.},
	author = {Ehsan, Upol and Liao, Q. Vera and Passi, Samir and Riedl, Mark O. and Daumé, Hal},
	month = apr,
	year = {2024},
	pages = {119:1--119:29},
}

@inproceedings{liu_agency_2025,
	address = {New York, NY, USA},
	series = {{HT} {Adjunct} '25},
	title = {Agency {Among} {Agents}: {Designing} with {Hypertextual} {Friction} in the {Algorithmic} {Web}},
	isbn = {979-8-4007-1533-4},
	shorttitle = {Agency {Among} {Agents}},
	url = {https://dl.acm.org/doi/10.1145/3720533.3750065},
	doi = {10.1145/3720533.3750065},
	urldate = {2026-02-06},
	booktitle = {Adjunct {Proceedings} of the 36th {ACM} {Conference} on {Hypertext} and {Social} {Media}},
	publisher = {Association for Computing Machinery},
	author = {Liu, Sophia and Almeda, Shm Garanganao},
	month = oct,
	year = {2025},
	pages = {30--34},
}

@inproceedings{ninomiya_determinants_2025,
	address = {New York, NY, USA},
	series = {{RecSys} '25},
	title = {Determinants of {Users}' {Chance}-{Seeking} {Behavior} in {Search}-{Based} {Recommendation}},
	isbn = {979-8-4007-1364-4},
	url = {https://dl.acm.org/doi/10.1145/3705328.3748019},
	doi = {10.1145/3705328.3748019},
	urldate = {2026-02-06},
	booktitle = {Proceedings of the {Nineteenth} {ACM} {Conference} on {Recommender} {Systems}},
	publisher = {Association for Computing Machinery},
	author = {Ninomiya, Yuki and Sone, Yutaro and Miwa, Kazuhisa and Sumi, Yuichiro and Nakanishi, Ryosuke and Mitsuda, Eiji and Sato, Koji and Odashima, Tadashi},
	month = sep,
	year = {2025},
	pages = {564--569},
}

@article{widener2025digital,
  title={Digital Media Trends: Social platforms are becoming a dominant force in media and entertainment},
  author={Widener, C and Arbanas, J and Van Dyke, D and Arkenberg, C and Matheson, B and Auxier, B},
  journal={Viitattu},
  volume={10},
  number={2025},
  pages={2025},
  year={2025}
}

@inproceedings{rhys_cox_beyond_2025,
	address = {New York, NY, USA},
	series = {C\&amp;{C} '25},
	title = {Beyond {Productivity}: {Rethinking} the {Impact} of {Creativity} {Support} {Tools}},
	isbn = {979-8-4007-1289-0},
	shorttitle = {Beyond {Productivity}},
	url = {https://dl.acm.org/doi/10.1145/3698061.3726924},
	doi = {10.1145/3698061.3726924},
	urldate = {2026-01-08},
	booktitle = {Proceedings of the 2025 {Conference} on {Creativity} and {Cognition}},
	publisher = {Association for Computing Machinery},
	author = {Rhys Cox, Samuel and Bøjer Djernæs, Helena and van Berkel, Niels},
	month = jun,
	year = {2025},
	pages = {735--749},
}

@inproceedings{pierce_tension_2021,
	address = {New York, NY, USA},
	series = {{CHI} '21},
	title = {In {Tension} with {Progression}: {Grasping} the {Frictional} {Tendencies} of {Speculative}, {Critical}, and other {Alternative} {Designs}},
	isbn = {978-1-4503-8096-6},
	shorttitle = {In {Tension} with {Progression}},
	url = {https://dl.acm.org/doi/10.1145/3411764.3445406},
	doi = {10.1145/3411764.3445406},
	urldate = {2026-02-05},
	booktitle = {Proceedings of the 2021 {CHI} {Conference} on {Human} {Factors} in {Computing} {Systems}},
	publisher = {Association for Computing Machinery},
	author = {Pierce, James},
	month = may,
	year = {2021},
	pages = {1--19},
}

@inproceedings{kreminski_reflective_2021,
	title = {Reflective {Creators}},
	url = {https://www.semanticscholar.org/paper/Reflective-Creators-Kreminski-Mateas/a7ea777833747312666b8d7af38b160535b2e2fa},
	urldate = {2026-01-08},
	author = {Kreminski, M. and Mateas, Michael},
	year = {2021},
      booktitle={ICCC},
}

@article{meinecke_towards_2022,
	title = {Towards {Enhancing} {Virtual} {Museums} by {Contextualizing} {Art} through {Interactive} {Visualizations}},
	volume = {15},
	issn = {1556-4673},
	url = {https://dl.acm.org/doi/10.1145/3527619},
	doi = {10.1145/3527619},
	number = {4},
	urldate = {2025-09-09},
	journal = {J. Comput. Cult. Herit.},
	author = {Meinecke, Christofer and Hall, Chris and Jänicke, Stefan},
	month = dec,
	year = {2022},
	pages = {62:1--62:26},
}

@article{avgousti_enhancing_2024,
	title = {Enhancing {Online} {Accessibility} of {Digitized} {Artifacts} from {Small} {Museum} {Collections} in {Cyprus}: {An} {Empirical} {Evaluation} of the {CyprusArk} {Solution}},
	volume = {17},
	issn = {1556-4673},
	shorttitle = {Enhancing {Online} {Accessibility} of {Digitized} {Artifacts} from {Small} {Museum} {Collections} in {Cyprus}},
	url = {https://dl.acm.org/doi/10.1145/3648229},
	doi = {10.1145/3648229},
	number = {3},
	urldate = {2025-09-09},
	journal = {J. Comput. Cult. Herit.},
	author = {Avgousti, A. and Papaioannou, G. and Hermon, S.},
	month = apr,
	year = {2024},
	pages = {34:1--34:24},
}

@inproceedings{kobeisse_moving_2023,
	address = {Tainan Taiwan},
	title = {Moving {Inside} the {Box}: {Interacting} with {Interpretation} of {Historical} {Artefacts} {Through} {Tangible} {Augmented} {Reality}},
	isbn = {979-8-4007-0205-1},
	shorttitle = {Moving {Inside} the {Box}},
	url = {https://dl.acm.org/doi/10.1145/3595916.3626408},
	doi = {10.1145/3595916.3626408},
	language = {en},
	urldate = {2025-09-09},
	booktitle = {{ACM} {Multimedia} {Asia} 2023},
	publisher = {ACM},
	author = {Kobeisse, Suzanne and Holmquist, Lars Erik},
	month = dec,
	year = {2023},
	pages = {1--7},
}

@article{murray_ava_2012,
	title = {{AVA}: {A} large-scale database for aesthetic visual analysis},
	shorttitle = {{AVA}},
	url = {http://ieeexplore.ieee.org/document/6247954/},
	doi = {10.1109/CVPR.2012.6247954},
	urldate = {2025-09-07},
	journal = {2012 IEEE Conference on Computer Vision and Pattern Recognition},
	author = {Murray, N. and Marchesotti, L. and Perronnin, F.},
	month = jun,
	year = {2012},
	note = {Conference Name: 2012 IEEE Conference on Computer Vision and Pattern Recognition (CVPR)
ISBN: 9781467312288 9781467312264 9781467312271
Place: Providence, RI
Publisher: IEEE},
	pages = {2408--2415},
}

@article{zhao_enhancing_2018,
	title = {Enhancing the {Appreciation} of {Traditional} {Chinese} {Painting} {Using} {Interactive} {Technology}},
	volume = {2},
	copyright = {http://creativecommons.org/licenses/by/3.0/},
	issn = {2414-4088},
	url = {https://www.mdpi.com/2414-4088/2/2/16},
	doi = {10.3390/mti2020016},
	language = {en},
	number = {2},
	urldate = {2025-09-07},
	journal = {Multimodal Technologies and Interaction},
	author = {Zhao, Shichao and Kirk, David and Bowen, Simon and Wright, Peter},
	month = jun,
	year = {2018},
	note = {Publisher: Multidisciplinary Digital Publishing Institute},
	pages = {16},
}

@inproceedings{kortbek_communicating_2008,
	address = {New York, NY, USA},
	series = {{NordiCHI} '08},
	title = {Communicating art through interactive technology: new approaches for interaction design in art museums},
	isbn = {978-1-59593-704-9},
	shorttitle = {Communicating art through interactive technology},
	url = {https://dl.acm.org/doi/10.1145/1463160.1463185},
	doi = {10.1145/1463160.1463185},
	urldate = {2025-09-07},
	booktitle = {Proceedings of the 5th {Nordic} conference on {Human}-computer interaction: building bridges},
	publisher = {Association for Computing Machinery},
	author = {Kortbek, Karen Johanne and Grønbæk, Kaj},
	month = oct,
	year = {2008},
	pages = {229--238},
}

@inproceedings{ciolfi_articulating_2016,
	address = {New York, NY, USA},
	series = {{CSCW} '16},
	title = {Articulating {Co}-{Design} in {Museums}: {Reflections} on {Two} {Participatory} {Processes}},
	isbn = {978-1-4503-3592-8},
	shorttitle = {Articulating {Co}-{Design} in {Museums}},
	url = {https://dl.acm.org/doi/10.1145/2818048.2819967},
	doi = {10.1145/2818048.2819967},
	urldate = {2025-09-07},
	booktitle = {Proceedings of the 19th {ACM} {Conference} on {Computer}-{Supported} {Cooperative} {Work} \& {Social} {Computing}},
	publisher = {Association for Computing Machinery},
	author = {Ciolfi, Luigina and Avram, Gabriela and Maye, Laura and Dulake, Nick and Marshall, Mark T. and van Dijk, Dick and McDermott, Fiona},
	month = feb,
	year = {2016},
	pages = {13--25},
}

@inproceedings{gorichanaz_engaging_2020,
	address = {New York, NY, USA},
	series = {{CHI} '20},
	title = {Engaging with {Public} {Art}: {An} {Exploration} of the {Design} {Space}},
	isbn = {978-1-4503-6708-0},
	shorttitle = {Engaging with {Public} {Art}},
	url = {https://dl.acm.org/doi/10.1145/3313831.3376640},
	doi = {10.1145/3313831.3376640},
	urldate = {2025-09-07},
	booktitle = {Proceedings of the 2020 {CHI} {Conference} on {Human} {Factors} in {Computing} {Systems}},
	publisher = {Association for Computing Machinery},
	author = {Gorichanaz, Tim},
	month = apr,
	year = {2020},
	pages = {1--14},
}

@article{zhang_inkthetics_2020,
	title = {Inkthetics: {A} {Comprehensive} {Computational} {Model} for {Aesthetic} {Evaluation} of {Chinese} {Ink} {Paintings}},
	volume = {8},
	issn = {2169-3536},
	shorttitle = {Inkthetics},
	url = {https://ieeexplore.ieee.org/abstract/document/9293299},
	doi = {10.1109/ACCESS.2020.3044573},
	urldate = {2025-09-07},
	journal = {IEEE Access},
	author = {Zhang, Jiajing and Miao, Yongwei and Zhang, Junsong and Yu, Jinhui},
	year = {2020},
	pages = {225857--225871},
}

@inproceedings{ryokai_artistic_2015,
	address = {New York, NY, USA},
	series = {{CHI} {EA} '15},
	title = {Artistic {Distance}: {Body} {Movements} as {Launching} {Points} for {Art} {Inquiry}},
	isbn = {978-1-4503-3146-3},
	shorttitle = {Artistic {Distance}},
	url = {https://dl.acm.org/doi/10.1145/2702613.2702958},
	doi = {10.1145/2702613.2702958},
	urldate = {2025-09-07},
	booktitle = {Proceedings of the 33rd {Annual} {ACM} {Conference} {Extended} {Abstracts} on {Human} {Factors} in {Computing} {Systems}},
	publisher = {Association for Computing Machinery},
	author = {Ryokai, Kimiko and Misra, Noriko and Hara, Yoshinori},
	month = apr,
	year = {2015},
	pages = {679--686},
}

@article{wakkary_situated_2007,
	title = {Situated play in a tangible interface and adaptive audio museum guide},
	volume = {11},
	issn = {1617-4909},
	url = {https://doi.org/10.1007/s00779-006-0101-8},
	doi = {10.1007/s00779-006-0101-8},
	number = {3},
	urldate = {2025-09-06},
	journal = {Personal Ubiquitous Comput.},
	author = {Wakkary, Ron and Hatala, Marek},
	month = feb,
	year = {2007},
	pages = {171--191},
}

@inproceedings{petrelli_phone_2018,
	address = {New York, NY, USA},
	series = {{CHI} '18},
	title = {Phone vs. {Tangible} in {Museums}: {A} {Comparative} {Study}},
	isbn = {978-1-4503-5620-6},
	shorttitle = {Phone vs. {Tangible} in {Museums}},
	url = {https://dl.acm.org/doi/10.1145/3173574.3173686},
	doi = {10.1145/3173574.3173686},
	urldate = {2025-09-06},
	booktitle = {Proceedings of the 2018 {CHI} {Conference} on {Human} {Factors} in {Computing} {Systems}},
	publisher = {Association for Computing Machinery},
	author = {Petrelli, Daniela and O'Brien, Sinead},
	month = apr,
	year = {2018},
	pages = {1--12},
}

@inproceedings{spence_seeing_2019,
	address = {New York, NY, USA},
	series = {{CHI} '19},
	title = {Seeing with {New} {Eyes}: {Designing} for {In}-the-{Wild} {Museum} {Gifting}},
	isbn = {978-1-4503-5970-2},
	shorttitle = {Seeing with {New} {Eyes}},
	url = {https://dl.acm.org/doi/10.1145/3290605.3300235},
	doi = {10.1145/3290605.3300235},
	urldate = {2025-09-06},
	booktitle = {Proceedings of the 2019 {CHI} {Conference} on {Human} {Factors} in {Computing} {Systems}},
	publisher = {Association for Computing Machinery},
	author = {Spence, Jocelyn and Bedwell, Benjamin and Coleman, Michelle and Benford, Steve and Koleva, Boriana N. and Adams, Matt and Row Farr, Ju and Tandavanitj, Nick and Løvlie, Anders Sundnes},
	month = may,
	year = {2019},
	pages = {1--13},
}

@inproceedings{weilenmann_instagram_2013,
	address = {New York, NY, USA},
	series = {{CHI} '13},
	title = {Instagram at the museum: communicating the museum experience through social photo sharing},
	isbn = {978-1-4503-1899-0},
	shorttitle = {Instagram at the museum},
	url = {https://dl.acm.org/doi/10.1145/2470654.2466243},
	doi = {10.1145/2470654.2466243},
	urldate = {2025-09-06},
	booktitle = {Proceedings of the {SIGCHI} {Conference} on {Human} {Factors} in {Computing} {Systems}},
	publisher = {Association for Computing Machinery},
	author = {Weilenmann, Alexandra and Hillman, Thomas and Jungselius, Beata},
	month = apr,
	year = {2013},
	pages = {1843--1852},
}

@article{levinson_refining_1989,
	title = {Refining {Art} {Historically}},
	volume = {47},
	issn = {0021-8529},
	url = {https://www.jstor.org/stable/431990},
	doi = {10.2307/431990},
	number = {1},
	urldate = {2025-09-06},
	journal = {The Journal of Aesthetics and Art Criticism},
	author = {Levinson, Jerrold},
	year = {1989},
	note = {Publisher: [Wiley, American Society for Aesthetics]},
	pages = {21--33},
}

@article{levinson_defining_1979,
	title = {{DEFINING} {ART} {HISTORICALLY}},
	volume = {19},
	issn = {0007-0904},
	url = {https://doi.org/10.1093/bjaesthetics/19.3.232},
	doi = {10.1093/bjaesthetics/19.3.232},
	number = {3},
	urldate = {2025-09-06},
	journal = {The British Journal of Aesthetics},
	author = {Levinson, Jerrold},
	month = jan,
	year = {1979},
	pages = {232--250},
}

@article{saleh_large-scale_2016,
	title = {Large-scale {Classification} of {Fine}-{Art} {Paintings}: {Learning} {The} {Right} {Metric} on {The} {Right} {Feature}},
	copyright = {Copyright (c) 2016 International Journal for Digital Art History},
	issn = {2363-5401},
	shorttitle = {Large-scale {Classification} of {Fine}-{Art} {Paintings}},
	url = {https://journals.ub.uni-heidelberg.de/index.php/dah/article/view/23376},
	doi = {10.11588/dah.2016.2.23376},
	language = {en},
	number = {2},
	urldate = {2025-09-06},
	journal = {International Journal for Digital Art History},
	author = {Saleh, Babak and Elgammal, Ahmed},
	month = oct,
	year = {2016},
}

@article{li_enhanced_2025,
	title = {Enhanced automated art curation using supervised modified {CNN} for art style classification},
	volume = {15},
	copyright = {2025 The Author(s)},
	issn = {2045-2322},
	url = {https://www.nature.com/articles/s41598-025-91671-z},
	doi = {10.1038/s41598-025-91671-z},
	language = {en},
	number = {1},
	urldate = {2025-09-06},
	journal = {Scientific Reports},
	author = {Li, Weiwei},
	month = mar,
	year = {2025},
	note = {Publisher: Nature Publishing Group},
	pages = {7319},
}

@article{luo_art_2025,
	title = {Art style classification via self-supervised dual-teacher knowledge distillation},
	volume = {174},
	issn = {1568-4946},
	url = {https://www.sciencedirect.com/science/article/pii/S1568494625002753},
	doi = {10.1016/j.asoc.2025.112964},
	urldate = {2025-09-06},
	journal = {Applied Soft Computing},
	author = {Luo, Mei and Liu, Li and Lu, Yue and Suen, Ching Y.},
	month = apr,
	year = {2025},
	pages = {112964},
}

@inproceedings{cameron_museum_2023,
	address = {New York, NY, USA},
	series = {Mindtrek '23},
	title = {The {Museum} is {Dreaming}: {Re}-{Imagining} the {Museum} through {Feminist} {Values} and {Data} {Practices} in {Design} {Fiction}},
	isbn = {979-8-4007-0874-9},
	shorttitle = {The {Museum} is {Dreaming}},
	url = {https://dl.acm.org/doi/10.1145/3616961.3616984},
	doi = {10.1145/3616961.3616984},
	urldate = {2025-09-05},
	booktitle = {Proceedings of the 26th {International} {Academic} {Mindtrek} {Conference}},
	publisher = {Association for Computing Machinery},
	author = {Cameron, Harriet R and Spors, Velvet},
	month = nov,
	year = {2023},
	pages = {182--194},
}

@inproceedings{almeda_creativity_2025,
	address = {New York, NY, USA},
	series = {{CHI} '25},
	title = {Creativity {Supportive} {Ecosystems}: {A} {Framework} for {Understanding} {Function} and {Disruption} in {Online} {Art} {Worlds}},
	isbn = {979-8-4007-1394-1},
	shorttitle = {Creativity {Supportive} {Ecosystems}},
	url = {https://dl.acm.org/doi/10.1145/3706598.3713734},
	doi = {10.1145/3706598.3713734},
	urldate = {2025-09-05},
	booktitle = {Proceedings of the 2025 {CHI} {Conference} on {Human} {Factors} in {Computing} {Systems}},
	publisher = {Association for Computing Machinery},
	author = {Almeda, Shm Garanganao and Kim, Joy O and Hartmann, Bjoern},
	month = apr,
	year = {2025},
	pages = {1--17},
}

@inproceedings{louie_expressive_2022,
	address = {New York, NY, USA},
	series = {{IUI} '22},
	title = {Expressive {Communication}: {Evaluating} {Developments} in {Generative} {Models} and {Steering} {Interfaces} for {Music} {Creation}},
	isbn = {978-1-4503-9144-3},
	shorttitle = {Expressive {Communication}},
	url = {https://dl.acm.org/doi/10.1145/3490099.3511159},
	doi = {10.1145/3490099.3511159},
	urldate = {2025-09-04},
	booktitle = {Proceedings of the 27th {International} {Conference} on {Intelligent} {User} {Interfaces}},
	publisher = {Association for Computing Machinery},
	author = {Louie, Ryan and Engel, Jesse and Huang, Cheng-Zhi Anna},
	month = mar,
	year = {2022},
	pages = {405--417},
}

@inproceedings{palani_evolving_2024,
	address = {New York, NY, USA},
	series = {C\&amp;{C} '24},
	title = {Evolving {Roles} and {Workflows} of {Creative} {Practitioners} in the {Age} of {Generative} {AI}},
	isbn = {9798400704857},
	url = {https://dl.acm.org/doi/10.1145/3635636.3656190},
	doi = {10.1145/3635636.3656190},
	urldate = {2025-04-10},
	booktitle = {Proceedings of the 16th {Conference} on {Creativity} \& {Cognition}},
	publisher = {Association for Computing Machinery},
	author = {Palani, Srishti and Ramos, Gonzalo},
	month = jun,
	year = {2024},
	pages = {170--184},
}

@inproceedings{li_beyond_2023,
	address = {New York, NY, USA},
	series = {{UIST} '23},
	title = {Beyond the {Artifact}: {Power} as a {Lens} for {Creativity} {Support} {Tools}},
	isbn = {9798400701320},
	shorttitle = {Beyond the {Artifact}},
	url = {https://dl.acm.org/doi/10.1145/3586183.3606831},
	doi = {10.1145/3586183.3606831},
	urldate = {2025-01-18},
	booktitle = {Proceedings of the 36th {Annual} {ACM} {Symposium} on {User} {Interface} {Software} and {Technology}},
	publisher = {Association for Computing Machinery},
	author = {Li, Jingyi and Rawn, Eric and Ritchie, Jacob and Tran O'Leary, Jasper and Follmer, Sean},
	month = oct,
	year = {2023},
	pages = {1--15},
}

@book{hooks_art_1995,
	title = {Art on {My} {Mind}: {Visual} {Politics}},
	isbn = {978-1-56584-263-2},
	shorttitle = {Art on {My} {Mind}},
	language = {en},
	publisher = {New Press},
	author = {Hooks, Bell},
	year = {1995},
}

@misc{carlini_extracting_2023,
	title = {Extracting {Training} {Data} from {Diffusion} {Models}},
	url = {http://arxiv.org/abs/2301.13188},
	doi = {10.48550/arXiv.2301.13188},
	urldate = {2024-07-15},
	publisher = {arXiv},
	author = {Carlini, Nicholas and Hayes, Jamie and Nasr, Milad and Jagielski, Matthew and Sehwag, Vikash and Tramèr, Florian and Balle, Borja and Ippolito, Daphne and Wallace, Eric},
	month = jan,
	year = {2023},
	note = {arXiv:2301.13188 [cs]},
}

@article{becker_art_1974,
	title = {Art {As} {Collective} {Action}},
	volume = {39},
	issn = {0003-1224},
	url = {https://www.jstor.org/stable/2094151},
	doi = {10.2307/2094151},
	number = {6},
	urldate = {2024-04-30},
	journal = {American Sociological Review},
	author = {Becker, Howard S.},
	year = {1974},
	note = {Publisher: [American Sociological Association, Sage Publications, Inc.]},
	pages = {767--776},
}

@inproceedings{chung_artist_2022,
	address = {New York, NY, USA},
	series = {{DIS} '22},
	title = {Artist {Support} {Networks}: {Implications} for {Future} {Creativity} {Support} {Tools}},
	isbn = {978-1-4503-9358-4},
	shorttitle = {Artist {Support} {Networks}},
	url = {https://dl.acm.org/doi/10.1145/3532106.3533505},
	doi = {10.1145/3532106.3533505},
	urldate = {2024-04-26},
	booktitle = {Proceedings of the 2022 {ACM} {Designing} {Interactive} {Systems} {Conference}},
	publisher = {Association for Computing Machinery},
	author = {Chung, John Joon Young and He, Shiqing and Adar, Eytan},
	month = jun,
	year = {2022},
	pages = {232--246},
}

@inproceedings{pancha_pinnerformer_2022,
	address = {New York, NY, USA},
	series = {{KDD} '22},
	title = {{PinnerFormer}: {Sequence} {Modeling} for {User} {Representation} at {Pinterest}},
	isbn = {978-1-4503-9385-0},
	shorttitle = {{PinnerFormer}},
	url = {https://dl.acm.org/doi/10.1145/3534678.3539156},
	doi = {10.1145/3534678.3539156},
	urldate = {2026-02-04},
	booktitle = {Proceedings of the 28th {ACM} {SIGKDD} {Conference} on {Knowledge} {Discovery} and {Data} {Mining}},
	publisher = {Association for Computing Machinery},
	author = {Pancha, Nikil and Zhai, Andrew and Leskovec, Jure and Rosenberg, Charles},
	month = aug,
	year = {2022},
	pages = {3702--3712},
}

@inproceedings{nakakoji_framework_1999,
	address = {New York, NY, USA},
	series = {C\&amp;{C} '99},
	title = {A framework that supports collective creativity in design using visual images},
	isbn = {978-1-58113-078-2},
	url = {https://dl.acm.org/doi/10.1145/317561.317590},
	doi = {10.1145/317561.317590},
	urldate = {2026-02-05},
	booktitle = {Proceedings of the 3rd conference on {Creativity} \& cognition},
	publisher = {Association for Computing Machinery},
	author = {Nakakoji, Kumiyo and Yamamoto, Yasuhiro and Ohira, Masao},
	month = oct,
	year = {1999},
	pages = {166--173},
}

@article{kato_power_2025,
	address = {Yokohama, Japan and online.},
	title = {Power, {Culture}, and {Sustainability} in {Creativity} {Support} {Tools}: {A} {Post}-growth {Perspective}},
	url = {https://junkato.jp/publications/chi2025-kato-postgrowth-cst.pdf},
	language = {en},
	journal = {In Hybrid Workshop: Advancing Post-growth HCI at CHI ’25},
	publisher = {ACM},
	author = {Kato, Jun and Yakura, Hiromu},
	month = apr,
	year = {2025},
	pages = {2},
}

@inproceedings{nakakoji_interaction_2002,
	address = {New York, NY, USA},
	series = {C\&amp;{C} '02},
	title = {Interaction design as a collective creative process},
	isbn = {978-1-58113-465-0},
	url = {https://dl.acm.org/doi/10.1145/581710.581727},
	doi = {10.1145/581710.581727},
	urldate = {2026-02-05},
	booktitle = {Proceedings of the 4th conference on {Creativity} \& cognition},
	publisher = {Association for Computing Machinery},
	author = {Nakakoji, Kumiyo and Yamamoto, Yasuhiro and Aoki, Atsushi},
	month = oct,
	year = {2002},
	pages = {103--110},
}

@article{glinka_critical-reflective_2023,
	title = {Critical-{Reflective} {Human}-{AI} {Collaboration}: {Exploring} {Computational} {Tools} for {Art} {Historical} {Image} {Retrieval}},
	volume = {7},
	issn = {2573-0142},
	shorttitle = {Critical-{Reflective} {Human}-{AI} {Collaboration}},
	url = {https://dl.acm.org/doi/10.1145/3610054},
	doi = {10.1145/3610054},
	language = {en},
	number = {CSCW2},
	urldate = {2026-02-05},
	journal = {Proceedings of the ACM on Human-Computer Interaction},
	author = {Glinka, Katrin and Müller-Birn, Claudia},
	month = sep,
	year = {2023},
	pages = {1--33},
}

@inproceedings{chen_pinfm_2025,
	address = {New York, NY, USA},
	series = {{RecSys} '25},
	title = {{PinFM}: {Foundation} {Model} for {User} {Activity} {Sequences} at a {Billion}-scale {Visual} {Discovery} {Platform}},
	isbn = {979-8-4007-1364-4},
	shorttitle = {{PinFM}},
	url = {https://dl.acm.org/doi/10.1145/3705328.3748050},
	doi = {10.1145/3705328.3748050},
	urldate = {2026-02-04},
	booktitle = {Proceedings of the {Nineteenth} {ACM} {Conference} on {Recommender} {Systems}},
	publisher = {Association for Computing Machinery},
	author = {Chen, Xiangyi and Rajesh, Kousik and Lawhon, Matthew and Wang, Zelun and Li, Hanyu and Li, Haomiao and Joshi, Saurabh Vishwas and Eksombatchai, Pong and Yang, Jaewon and Hsu, Yi-Ping and Xu, Jiajing and Rosenberg, Charles},
	month = sep,
	year = {2025},
	pages = {381--390},
}

@inproceedings{chung_intersection_2021,
	address = {New York, NY, USA},
	series = {{DIS} '21},
	title = {The {Intersection} of {Users}, {Roles}, {Interactions}, and {Technologies} in {Creativity} {Support} {Tools}},
	isbn = {978-1-4503-8476-6},
	url = {https://dl.acm.org/doi/10.1145/3461778.3462050},
	doi = {10.1145/3461778.3462050},
	urldate = {2024-04-26},
	booktitle = {Proceedings of the 2021 {ACM} {Designing} {Interactive} {Systems} {Conference}},
	publisher = {Association for Computing Machinery},
	author = {Chung, John Joon Young and He, Shiqing and Adar, Eytan},
	month = jun,
	year = {2021},
	pages = {1817--1833},
}

@inproceedings{frich_mapping_2019,
	address = {New York, NY, USA},
	series = {{CHI} '19},
	title = {Mapping the {Landscape} of {Creativity} {Support} {Tools} in {HCI}},
	isbn = {978-1-4503-5970-2},
	url = {https://dl.acm.org/doi/10.1145/3290605.3300619},
	doi = {10.1145/3290605.3300619},
	urldate = {2024-04-26},
	booktitle = {Proceedings of the 2019 {CHI} {Conference} on {Human} {Factors} in {Computing} {Systems}},
	publisher = {Association for Computing Machinery},
	author = {Frich, Jonas and MacDonald Vermeulen, Lindsay and Remy, Christian and Biskjaer, Michael Mose and Dalsgaard, Peter},
	month = may,
	year = {2019},
	pages = {1--18},
}

@inproceedings{frich_twenty_2018,
	address = {New York, NY, USA},
	series = {{DIS} '18},
	title = {Twenty {Years} of {Creativity} {Research} in {Human}-{Computer} {Interaction}: {Current} {State} and {Future} {Directions}},
	isbn = {978-1-4503-5198-0},
	shorttitle = {Twenty {Years} of {Creativity} {Research} in {Human}-{Computer} {Interaction}},
	url = {https://dl.acm.org/doi/10.1145/3196709.3196732},
	doi = {10.1145/3196709.3196732},
	urldate = {2024-04-26},
	booktitle = {Proceedings of the 2018 {Designing} {Interactive} {Systems} {Conference}},
	publisher = {Association for Computing Machinery},
	author = {Frich, Jonas and Mose Biskjaer, Michael and Dalsgaard, Peter},
	month = jun,
	year = {2018},
	pages = {1235--1257},
}

@article{von_davier_machine_2024,
    title = {A {Machine} {Walks} into an {Exhibit}: {A} {Technical} {Analysis} of {Art} {Curation}},
    volume = {13},
    copyright = {http://creativecommons.org/licenses/by/3.0/},
    issn = {2076-0752},
    shorttitle = {A {Machine} {Walks} into an {Exhibit}},
    url = {https://www.mdpi.com/2076-0752/13/5/138},
    doi = {10.3390/arts13050138},
    language = {en},
    number = {5},
    urldate = {2025-09-06},
    journal = {Arts},
    author = {von Davier, Thomas Şerban and Herman, Laura M. and Moruzzi, Caterina},
    month = oct,
    year = {2024},
    note = {Publisher: Multidisciplinary Digital Publishing Institute},
    pages = {138},
}

@inproceedings{bardzell_interaction_2008,
	address = {New York, NY, USA},
	series = {{CHI} {EA} '08},
	title = {Interaction criticism: a proposal and framework for a new discipline of hci},
	isbn = {978-1-60558-012-8},
	shorttitle = {Interaction criticism},
	url = {https://dl.acm.org/doi/10.1145/1358628.1358703},
	doi = {10.1145/1358628.1358703},
	urldate = {2026-01-18},
	booktitle = {{CHI} '08 {Extended} {Abstracts} on {Human} {Factors} in {Computing} {Systems}},
	publisher = {Association for Computing Machinery},
	author = {Bardzell, Jeffrey and Bardzell, Shaowen},
	month = apr,
	year = {2008},
	pages = {2463--2472},
}

@inproceedings{bar_classification_2015,
    address = {Cham},
    title = {Classification of {Artistic} {Styles} {Using} {Binarized} {Features} {Derived} from a {Deep} {Neural} {Network}},
    isbn = {978-3-319-16178-5},
    doi = {10.1007/978-3-319-16178-5_5},
    language = {en},
    booktitle = {Computer {Vision} - {ECCV} 2014 {Workshops}},
    publisher = {Springer International Publishing},
    author = {Bar, Yaniv and Levy, Noga and Wolf, Lior},
    editor = {Agapito, Lourdes and Bronstein, Michael M. and Rother, Carsten},
    year = {2015},
    pages = {71--84},
}

@inproceedings{yilma_elements_2023,
    address = {New York, NY, USA},
    series = {{CHI} '23},
    title = {The {Elements} of {Visual} {Art} {Recommendation}: {Learning} {Latent} {Semantic} {Representations} of {Paintings}},
    isbn = {978-1-4503-9421-5},
    shorttitle = {The {Elements} of {Visual} {Art} {Recommendation}},
    url = {https://dl.acm.org/doi/10.1145/3544548.3581477},
    doi = {10.1145/3544548.3581477},
    urldate = {2025-09-09},
    booktitle = {Proceedings of the 2023 {CHI} {Conference} on {Human} {Factors} in {Computing} {Systems}},
    publisher = {Association for Computing Machinery},
    author = {Yilma, Bereket A. and Leiva, Luis A.},
    month = apr,
    year = {2023},
    pages = {1--17},
}

@inproceedings{sheahan_designing_2024,
	address = {New York, NY, USA},
	series = {{NordiCHI} '24 {Adjunct}},
	title = {Designing with {Friction}: {Inverting} {Notions} of {Seamless} {Technology}},
	isbn = {979-8-4007-0965-4},
	shorttitle = {Designing with {Friction}},
	url = {https://dl.acm.org/doi/10.1145/3677045.3685504},
	doi = {10.1145/3677045.3685504},
	urldate = {2026-05-13},
	booktitle = {Adjunct {Proceedings} of the 2024 {Nordic} {Conference} on {Human}-{Computer} {Interaction}},
	publisher = {Association for Computing Machinery},
	author = {Sheahan, Jacob and Chatting, David and Collins, Robert and Bley, Jessica and Eriksson, Alexander and Taylor, Nick and Rozendaal, Marco C.},
	month = oct,
	year = {2024},
	pages = {1--4},
}

@inproceedings{bederson1994pad,
author = {Bederson, Benjamin B. and Hollan, James D.},
title = {Pad++: a zooming graphical interface for exploring alternate interface physics},
year = {1994},
isbn = {0897916573},
publisher = {Association for Computing Machinery},
address = {New York, NY, USA},
url = {https://doi.org/10.1145/192426.192435},
doi = {10.1145/192426.192435},
booktitle = {Proceedings of the 7th Annual ACM Symposium on User Interface Software and Technology},
pages = {17–26},
numpages = {10},
location = {Marina del Rey, California, USA},
series = {UIST '94}
}

@inproceedings{lamping1995focus,
author = {Lamping, John and Rao, Ramana and Pirolli, Peter},
title = {A focus+context technique based on hyperbolic geometry for visualizing large hierarchies},
year = {1995},
isbn = {0201847051},
publisher = {ACM Press/Addison-Wesley Publishing Co.},
address = {USA},
url = {https://doi.org/10.1145/223904.223956},
doi = {10.1145/223904.223956},
booktitle = {Proceedings of the SIGCHI Conference on Human Factors in Computing Systems},
pages = {401–408},
numpages = {8},
location = {Denver, Colorado, USA},
series = {CHI '95}
}

@article{yi2005dust,
author = {Yi, Ji Soo and Melton, Rachel and Stasko, John and Jacko, Julie A.},
title = {Dust \& magnet: multivariate information visualization using a magnet metaphor},
year = {2005},
issue_date = {October 2005},
publisher = {Palgrave Macmillan},
volume = {4},
number = {4},
issn = {1473-8716},
url = {https://doi.org/10.1057/palgrave.ivs.9500099},
doi = {10.1057/palgrave.ivs.9500099},
journal = {Information Visualization},
month = oct,
pages = {239–256},
numpages = {18}
}

@inproceedings{fass2000picturepiper,
author = {Fass, Adam M. and Bier, Eric A. and Adar, Eyton},
title = {PicturePiper: using a re-configurable pipeline to find images on the Web},
year = {2000},
isbn = {1581132123},
publisher = {Association for Computing Machinery},
address = {New York, NY, USA},
url = {https://doi.org/10.1145/354401.354411},
doi = {10.1145/354401.354411},
booktitle = {Proceedings of the 13th Annual ACM Symposium on User Interface Software and Technology},
pages = {51–62},
numpages = {12},
location = {San Diego, California, USA},
series = {UIST '00}
}

@inproceedings{suh2023sensecape,
author = {Suh, Sangho and Min, Bryan and Palani, Srishti and Xia, Haijun},
title = {Sensecape: Enabling Multilevel Exploration and Sensemaking with Large Language Models},
year = {2023},
isbn = {9798400701320},
publisher = {Association for Computing Machinery},
address = {New York, NY, USA},
url = {https://doi.org/10.1145/3586183.3606756},
doi = {10.1145/3586183.3606756},
booktitle = {Proceedings of the 36th Annual ACM Symposium on User Interface Software and Technology},
articleno = {1},
numpages = {18},
location = {San Francisco, CA, USA},
series = {UIST '23}
}

@inproceedings{luminate2024suh,
author = {Suh, Sangho and Chen, Meng and Min, Bryan and Li, Toby Jia-Jun and Xia, Haijun},
title = {Luminate: Structured Generation and Exploration of Design Space with Large Language Models for Human-AI Co-Creation},
year = {2024},
isbn = {9798400703300},
publisher = {Association for Computing Machinery},
address = {New York, NY, USA},
url = {https://doi.org/10.1145/3613904.3642400},
doi = {10.1145/3613904.3642400},
booktitle = {Proceedings of the 2024 CHI Conference on Human Factors in Computing Systems},
articleno = {644},
numpages = {26},
location = {Honolulu, HI, USA},
series = {CHI '24}
}

@inproceedings{chung2024patchview,
author = {Chung, John Joon Young and Kreminski, Max},
title = {Patchview: LLM-powered Worldbuilding with Generative Dust and Magnet Visualization},
year = {2024},
isbn = {9798400706288},
publisher = {Association for Computing Machinery},
address = {New York, NY, USA},
url = {https://doi.org/10.1145/3654777.3676352},
doi = {10.1145/3654777.3676352},
booktitle = {Proceedings of the 37th Annual ACM Symposium on User Interface Software and Technology},
articleno = {77},
numpages = {19},
location = {Pittsburgh, PA, USA},
series = {UIST '24}
}

@inproceedings{isbister_sensual_2006,
    address = {New York, NY, USA},
    series = {{CHI} '06},
    title = {The sensual evaluation instrument: developing an affective evaluation tool},
    isbn = {978-1-59593-372-0},
    shorttitle = {The sensual evaluation instrument},
    url = {https://dl.acm.org/doi/10.1145/1124772.1124946},
    doi = {10.1145/1124772.1124946},
    urldate = {2025-09-10},
    booktitle = {Proceedings of the {SIGCHI} {Conference} on {Human} {Factors} in {Computing} {Systems}},
    publisher = {Association for Computing Machinery},
    author = {Isbister, Katherine and Höök, Kristina and Sharp, Michael and Laaksolahti, Jarmo},
    month = apr,
    year = {2006},
    pages = {1163--1172},
}

@inproceedings{nelson_curious_2018,
    address = {New York, NY, USA},
    series = {{FDG} '18},
    title = {Curious users of casual creators},
    isbn = {978-1-4503-6571-0},
    url = {https://dl.acm.org/doi/10.1145/3235765.3235826},
    doi = {10.1145/3235765.3235826},
    urldate = {2025-09-10},
    booktitle = {Proceedings of the 13th {International} {Conference} on the {Foundations} of {Digital} {Games}},
    publisher = {Association for Computing Machinery},
    author = {Nelson, Mark J. and Gaudl, Swen E. and Colton, Simon and Deterding, Sebastian},
    month = aug,
    year = {2018},
    pages = {1--6},
}

@article{tuscher_nodes_2025,
    title = {Nodes, {Edges}, and {Artistic} {Wedges}: {A} {Survey} on {Network} {Visualization} in {Art} {History}},
    volume = {44},
    copyright = {© 2025 The Author(s). Computer Graphics Forum published by Eurographics - The European Association for Computer Graphics and John Wiley \& Sons Ltd.},
    issn = {1467-8659},
    shorttitle = {Nodes, {Edges}, and {Artistic} {Wedges}},
    url = {https://onlinelibrary.wiley.com/doi/abs/10.1111/cgf.70154},
    doi = {10.1111/cgf.70154},
    language = {en},
    number = {3},
    urldate = {2025-09-11},
    journal = {Computer Graphics Forum},
    author = {Tuscher, Michaela and Filipov, Velitchko and Kamencek, Teresa and Rosenberg, Raphael and Miksch, Silvia},
    year = {2025},
    note = {\_eprint: https://onlinelibrary.wiley.com/doi/pdf/10.1111/cgf.70154},
    pages = {e70154},
}

@misc{schuhmann_christophschuhmannimproved-aesthetic-predictor_2025,
    title = {christophschuhmann/improved-aesthetic-predictor},
    copyright = {Apache-2.0},
    url = {https://github.com/christophschuhmann/improved-aesthetic-predictor},
    urldate = {2025-09-11},
    author = {Schuhmann, Christoph},
    month = sep,
    year = {2025},
    note = {original-date: 2022-06-25T20:57:49Z},
}

@inproceedings{srinivasan_see_2024,
    address = {New York, NY, USA},
    series = {{FAccT} '24},
    title = {To {See} or {Not} to {See}: {Understanding} the {Tensions} of {Algorithmic} {Curation} for {Visual} {Arts}},
    isbn = {979-8-4007-0450-5},
    shorttitle = {To {See} or {Not} to {See}},
    url = {https://dl.acm.org/doi/10.1145/3630106.3658917},
    doi = {10.1145/3630106.3658917},
    urldate = {2025-09-10},
    booktitle = {Proceedings of the 2024 {ACM} {Conference} on {Fairness}, {Accountability}, and {Transparency}},
    publisher = {Association for Computing Machinery},
    author = {Srinivasan, Ramya},
    month = jun,
    year = {2024},
    pages = {444--455},
}

@misc{wikiartorg_hugganwikiart_2025,
    title = {huggan/wikiart · {Datasets} at {Hugging} {Face}},
    url = {https://huggingface.co/datasets/huggan/wikiart},
    urldate = {2025-09-11},
    publisher = {Hugging Face},
    author = {wikiart.org},
    month = jun,
    year = {2025},
}

@article{mcinnes_umap_2018,
    title = {{UMAP}: {Uniform} {Manifold} {Approximation} and {Projection}},
    volume = {3},
    issn = {2475-9066},
    shorttitle = {{UMAP}},
    url = {https://joss.theoj.org/papers/10.21105/joss.00861},
    doi = {10.21105/joss.00861},
    language = {en},
    number = {29},
    urldate = {2025-09-11},
    journal = {Journal of Open Source Software},
    author = {McInnes, Leland and Healy, John and Saul, Nathaniel and Großberger, Lukas},
    month = sep,
    year = {2018},
    pages = {861},
}

@book{saldana_coding_2015,
    title = {The {Coding} {Manual} for {Qualitative} {Researchers}},
    isbn = {978-1-4739-4358-2},
    language = {en},
    publisher = {SAGE},
    author = {Saldana, Johnny},
    month = nov,
    year = {2015},
    note = {Google-Books-ID: jh1iCgAAQBAJ},
}

@article{braun_reflecting_2019,
    title = {Reflecting on reflexive thematic analysis},
    volume = {11},
    issn = {2159-676X},
    url = {https://doi.org/10.1080/2159676X.2019.1628806},
    doi = {10.1080/2159676X.2019.1628806},
    number = {4},
    urldate = {2025-09-11},
    journal = {Qualitative Research in Sport, Exercise and Health},
    author = {Braun, Virginia and Clarke, Victoria},
    month = aug,
    year = {2019},
    note = {Publisher: Routledge
\_eprint: https://doi.org/10.1080/2159676X.2019.1628806},
    pages = {589--597},
}

@inproceedings{compositiontools,
  title={Design principles for tools to support creative thinking},
  author={Resnick, Mitchel and Myers, Brad and Nakakoji, Kumiyo and Shneiderman, Ben and Pausch, Randy and Selker, Ted and Eisenberg, Mike},
  year={2005},
  booktitle={Report of Workshop on Creativity Support Tools}
}

@inproceedings{CasualCreators,
  title={Casual creators},
  author={Compton, Kate and Mateas, Michael},
  booktitle={International Conference on Computational Creativity},
  pages={228--235},
  year={2015}
}
